\documentclass[11pt]{article}
\usepackage{amsmath, amssymb, graphics}
\usepackage[nosort]{cite}
\newcommand{\mathsym}[1]{{}}
\usepackage{verbatim}
\usepackage{amsfonts,euscript,amssymb,stmaryrd,braket}
\usepackage{graphics,tikz}
\usetikzlibrary{shapes.misc,arrows,decorations.markings,patterns,calc,shapes.geometric}
\usepackage{caption}
\usepackage{subcaption}
\usepackage{slashed}
\definecolor{hyperref}{RGB}{026,028,185}
\usepackage[bookmarks=true,colorlinks=true,linkcolor=hyperref,citecolor=hyperref,urlcolor=hyperref,bookmarksnumbered]{hyperref}
 \usepackage{booktabs}
 \usepackage{multirow}
\usepackage{tabularx}
\usepackage{longtable}
\usepackage{bbold,bbding}
\usepackage{amsthm} 
\usepackage{esint}
\newcommand{\bal}{\begin{equation}\begin{aligned}}
\newcommand{\eal}{\end{aligned} \end{equation}}
\def\id{\protect{{1 \kern-.28em {\rm l}}}}
\def\dd{\text{dDisc}}
  
\makeatletter
\renewcommand\section{\@startsection {section}{1}{\z@}%
                                   {-3.5ex \@plus -1ex \@minus -.2ex}%
                                   {2.3ex \@plus.2ex}%
                                   {\normalfont\large\bfseries}}
\renewcommand\subsection{\@startsection{subsection}{2}{\z@}%
                                   {-3.25ex\@plus -1ex \@minus -.2ex}%
                                   {1.5ex \@plus .2ex}%
                                   {\normalfont\normalsize\bfseries}}

\makeatother

\numberwithin{equation}{section}

\usepackage[utf8]{inputenc}
\usepackage{epsfig}
\usepackage{graphicx}
\usepackage[multiple]{footmisc}
\usepackage{amssymb,amsmath}
\usepackage{fancybox,framed,tikz}
\usetikzlibrary{decorations.pathmorphing,patterns}
\usepackage{dsfont}
\usepackage{mathtools}
\usepackage{braket}
\usepackage{slashed}
\usepackage{rotating}
\usepackage{bbold,amsfonts}
\usepackage{multirow}
\usepackage{verbatim}
\usepackage{upgreek}

\tikzset{cross/.style={cross out, draw=black, minimum size=2*(#1-\pgflinewidth), inner sep=0pt, outer sep=0pt},
cross/.default={1pt}}

\newcommand{\be}{\begin{equation}}
\newcommand{\ee}{\end{equation}}

\usepackage{color}

\definecolor{mypink1}{rgb}{0.958, 0.188, 0.478}

\newcommand{\ba}{\begin{eqnarray}}
\newcommand{\ea}{\end{eqnarray}}

\def\G{\mathcal{G}}

\hypersetup{}
 
\tikzset{Witten diagram/.style={execute at begin picture={%
\draw[blue ,fill=blue!05] curve[radius=\pgfkeysvalueof{/tikz/Witten/radius}];
\path node (X){\phantom{X}};
},baseline={(X.base)}},vertex/.style={curve,fill,inner sep=1.414pt,node
contents={}},
Witten/.cd,radius/.initial=1.414cm}

\begin{document}
\renewcommand{\thefootnote}{\arabic{footnote}}

\overfullrule=0pt
\parskip=2pt
\parindent=12pt
\headheight=0in \headsep=0in \topmargin=0in \oddsidemargin=0in

\vspace{ -3cm} \thispagestyle{empty} \vspace{-1cm}
\begin{flushright} 
\footnotesize
{HU-EP-24/19}
 
\end{flushright}%

\begin{center}
\vspace{1.2cm}
{\Large\bf \mathversion{bold}
{Dispersion relation from Lorentzian inversion in 1d CFT}
}
 
\author{ABC\thanks{XYZ} \and DEF\thanks{UVW} \and GHI\thanks{XYZ}}
 \vspace{0.8cm} {
  Davide Bonomi$^{a,b}$\footnote{{\tt davide.bonomi@city.ac.uk}}
and  Valentina~Forini$^{b}$\footnote{{\tt forini@physik.hu-berlin.de }}
 }
 \vskip  0.5cm

\small
{\em

$^{a}$   Department of Mathematics, City, University of London,\\
Northampton Square, EC1V 0HB London, United Kingdom
\vskip 0.02cm
$^{b}$  
Institut f\"ur Physik, Humboldt-Universit\"at zu Berlin, \\Zum Gro\ss en Windkanal 2, 12489 Berlin, Germany\\

}
\normalsize

\end{center}

\vspace{0.3cm}
\begin{abstract} 
Starting from the Lorentzian inversion formula, we derive a dispersion relation which computes a four-point function in 1d CFTs as an integral over its double discontinuity.  The  crossing symmetric kernel of the integral  is given explicitly for the case of identical operators with integer or half-integer scaling dimension. This derivation complements the one that uses analytic functionals. We use the dispersion relation to evaluate holographic correlators defined on the half-BPS Wilson line of planar $\mathcal{N}=4$ super Yang-Mills, reproducing results up to fourth order in an expansion at large t'Hooft coupling.  
\end{abstract}
\tableofcontents

 \section{Discussion}

A crucial tool for the analytical approach to the conformal bootstrap~\cite{Bissi:2022mrs,Hartman:2022zik} is the Lorentzian inversion formula of~\cite{Caron-Huot:2017vep},  which allows to obtain the CFT data of a given four-point correlator from its (double) discontinuity.   The latter can then be taken as the starting point to reconstruct the full correlator, in which case one talks about a dispersion relation~\cite{Carmi:2019cub}, in analogy with the ones arising in the context of the original S-matrix bootstrap~\cite{Eden:1966dnq}.

For CFT correlators restricted to the line, dispersion relations have been derived  in~\cite{Paulos:2020zxx} exploiting the formalism of analytic functionals~\cite{Mazac:2016qev,Mazac:2018mdx,Mazac:2018ycv}. In~\cite{Paulos:2020zxx},  the action of a class of  ``master'' functionals  on the crossing equation for the correlator has been shown to generate a family of dispersive sum rules~\footnote{ We call a sum rule “dispersive” if it has double zeros at the values of the dimensions of two-particle operators.}, which in turn can be reinterpreted as dispersion relations.  The   functional kernels can be computed case by case numerically or, in the case of correlators of operators with integer or half-integer dimension, analytically  using an Ansatz to solve the corresponding master functional equations.     

In this paper we  complement the analytical treatment of conformal correlators in a CFT$_1$,  deriving the  dispersion relation  directly from the Lorentzian inversion formula, which in the one-dimensional case has been obtained in~\cite{Simmons-Duffin:2017nub,Mazac:2018qmi}.

The dispersion relation, discussed in Section~\ref{sec:kernel}, reads
\begin{equation}\label{dispersion}
    \mathcal{G}(z) = \int_{0}^{1} dw \, w^{-2} \ \dd\,[\mathcal{G}(w)] K_{\Delta_\phi}(z,w)\,.
\end{equation}
The  input of the formula  is the double discontinuity $\dd\,\G(z)$,  
defined below in~\eqref{ddisc}, which 
in the case of identical bosons can be expressed in terms of the OPE data as
\begin{equation}
    \dd \big[\mathcal{G}(z)]= \sum_\Delta  2\sin^{2}{\frac{\pi}{2} (\Delta-2\Delta_{\phi})} a_\Delta\frac{z^{2\Delta_{\phi}}}{(1-z)^{2\Delta_{\phi}}} G_{\Delta}(1-z) \, ,
\end{equation}
or,  in the fermionic case,
\begin{equation}
    \dd \big[\mathcal{G}(z)]= \sum_\Delta  2\cos^{2}{\frac{\pi}{2} (\Delta-2\Delta_{\phi})} a_\Delta\frac{z^{2\Delta_{\phi}}}{(1-z)^{2\Delta_{\phi}}} G_{\Delta}(1-z) \, .
\end{equation}
Both expressions have double zeros at the dimensions of two-particle operators~\footnote{These are operators, of symbolic form $\phi\, \square^n \phi$, exchanged in the generalized free field theory correlator. In the literature they are often dubbed, borrowing from their higher-dimensional counterpart, as double-twist or double-trace operators.}, which are $\Delta = 2\Delta_\phi +2n$ in the bosonic case and  $\Delta = 2\Delta_\phi +2n+1$ in the fermionic one. This property allows to derive dispersive sum rules for the OPE data from the dispersion relation, as explained in \cite{Mazac:2018qmi,Paulos:2020zxx}.

Below, we work out explicitly the  kernel $K_{\Delta_\phi}(z,w)$ in~\eqref{dispersion} for correlators of identical operators with integer or half-integer dimension $\Delta_\phi$. In the case of Regge-(super)bounded~\footnote{See~\eqref{superbounded} and~\eqref{bounded} below for the definition of Regge-superbounded and Regge-bounded correlators. } (bosonic) fermionic correlators, it reads
\begin{align}\label{Kfinalin}
       K_{\Delta_\phi}(z,w) &= \frac{w\, z^2 (w-2) \log (1-w)}{\pi ^2 (w-z) (w+z-wz)}-\frac{z\,w^2 (z-2)  \log (1-z)}{\pi ^2 (w-z) (w+z-wz)}\\\nonumber
        &\pm \frac{z^2 }{\pi ^2}  \Big[\,\textstyle\!\!\log(1\!-\!w)\frac{(1-2w) w^{2-2 \Delta_\phi }}{(w-1) w z^2+z-1}    + \frac{\log (1-z) }{z} \textstyle \frac{w^{2-2 \Delta_\phi }}{w z-1}  + \log (z)  \frac{(1-2w) w^{2-2 \Delta_\phi }}{(w-1) w z^2+z-1}+(\textstyle w\! \rightarrow \!\frac{w}{w-1} )\,\Big]\\\nonumber
        & \textstyle +\sum\limits_{m=0}^{2\Delta_\phi - 2} \sum\limits_{n=0}^{2\Delta_\phi - 4}  (\alpha_{m,n}+\beta_{m,n}\log(1-w)) w^{m+2-2\Delta_\phi} \,\mathcal{C}^n \left[\frac{2}{\pi^2} \left(\frac{z^2 \log (z)}{1-z}+z \log (1-z)\right)\right] \, .
\end{align}
where the plus sign is for bosons, the minus for fermions. The coefficients $\alpha_{m,n}, \beta_{m,n}$  can be determined by solving a system of equations
\begin{eqnarray}\nonumber
         &&\!\!\!\!\!\textstyle\sum\limits_{m=0}^{2\Delta_\phi - 2} \sum\limits_{n=0}^{2\Delta_\phi - 4} (\alpha_{m,n}+\beta_{m,n}\log(1-w)) w^{m+2-2\Delta_\phi} \Big(\mathcal{C}^n \Big[\frac{2}{\pi^2} \Big(\frac{z^2 \log (z)}{1-z}+z \log (1-z)\Big)\Big] \!\!-\text{crossing} \Big) = \\\label{1.5}
        & &= \frac{z^{2\Delta_\phi}}{(1-z)^{2\Delta_\phi}} \left( K^{\text{discrete}}(1-z,w) + K_{\Delta_\phi}^{p}(1-z,w) \right) -K^{\text{discrete}}(z,w) - K_{\Delta_\phi}^{p}(z,w) \, ,
\end{eqnarray}
where $K^{\text{discrete}}(z,w)$ and $ K_{\Delta_\phi}^{p}(z,w)$ are given explicitly in \eqref{Kdiscrete} and \eqref{Kpcomp} respectively, and $\mathcal{C}$ is the Casimir operator $\mathcal{C} =z^2 (1-z) \partial^2 -z^2 \partial$. In~\eqref{1.5}, with  ``crossing'' we indicated the expression $ \textstyle\mathcal{C}^n  [\frac{2}{\pi^2}  (\frac{z^2 \log (z)}{1-z}+z \log (1-z) ) ]$ evaluated at $z\rightarrow 1-z$ and multiplied by $\textstyle\frac{z^{2\Delta_\phi}}{(1-z)^{2\Delta_\phi}}$, see discussion around the the crossing equation~\eqref{crossing} below. 
The kernel~\eqref{Kfinalin} is positive, as one can easily check by plotting it for specific values of integer and semi-integer $\Delta_\phi$.  As discussed below, for Regge-bounded bosonic correlators, the dispersion relation acquires extra contributions and requires a regularisation of the correlator. This reads
\begin{equation}
 \begin{split}
        \mathcal{G}^{\text{reg}} (z) &=  \int_{0}^{1} \frac{dw}{w^2} K^{\text{bd}}_{\Delta_\phi}(z,w) \, \text{dDisc}\mathcal{G}^{\text{reg}}(w) + \\
& +\lim _{\rho \rightarrow 0} \int_{C_\rho^{+}} \frac{dw}{2w^2}  K^{\text{bd}}_{\Delta_\phi}(z,w) \mathcal{G}^{\text{reg}}(w)+\lim _{\rho \rightarrow 0} \int_{C_\rho^{-}} \frac{dw}{2w^2} K^{\text{bd}}_{\Delta_\phi}(z,w) \mathcal{G}^{\text{reg}}(w) \, ,
 \end{split}
\end{equation}
where $C_\rho^{\pm}$ are semicircular contours going above and below $w=1$ in the complex plane.
The kernel $ K^{\text{bd}}_{\Delta_\phi}(z,w)$ for Regge-bounded  bosonic correlators is given in \eqref{Kimpr}. From its explicit expression, one can see that the kernel has a pole in $w=1$. For this reason it is necessary to introduce a regularized correlator $\mathcal{G}^{\text{reg}}(z)$ by performing a subtraction, which in general depends on the CFT data of low-dimensional operators with $\Delta < 2\Delta_\phi$.  
A dispersion relation can also be constructed for correlators that diverge in the Regge limit, and the corresponding kernel $K^{\text{unbd}}(z,w)$ is given in~\eqref{Kunbounded}.

Notice that it is not possible to derive the dispersion relation from a diagonal limit ($z=\bar z$) of the higher-dimensional expression in~\cite{Carmi:2019cub}, as the one-dimensional case is  intrinsically different.  Indeed, the higher-dimensional inversion formula~\cite{Caron-Huot:2017vep} does not converge for $d=1$, or for scalar operators. Moreover, in the one-dimensionial case the four-point function can be expanded in a
complete set of conformal partial waves which includes contributions from both the principal and the discrete series of $SL(2,\mathbb{R})$~\cite{Harish-Chandra}, the latter being absent in $d>1$. The two contributions give rise to two distinct inversion formulae~\cite{Simmons-Duffin:2017nub,Mazac:2018qmi}. These differences at the level of the inversion formula imply the necessity of deriving a dispersion formula directly in $d=1$.

At variance with the higher-dimensional case~\cite{Carmi:2019cub}, the kernel~\eqref{Kfinalin} depends explicitly on the dimensions $\Delta_{\phi}$ of the external operators and it is manifestly crossing symmetric. 
As pointed out in \cite{Paulos:2020zxx}, this fact implies the equivalence between the dispersion relation and the Polyakov bootstrap~\cite{Polyakov:1974gs, Gopakumar:2016wkt,Ghosh:2021ruh,Kaviraj:2021cvq,Ghosh:2023wjn,Ferrero:2019luz}. 
This is the idea of replacing the conformal block expansion with a similar expansion in terms of crossing symmetric Polyakov blocks $\mathcal{P}^{\Delta_\phi}_{\Delta}(z)$ such that
\be
 \mathcal{G}(z) = \sum_{\Delta} a_{\Delta}  G_{\Delta} (z) =  \sum_{\Delta} a_{\Delta} \mathcal{P}^{\Delta_\phi}_{\Delta}(z) \,.
\ee
The dispersion relation can be used to obtain an integral representation of the Polyakov blocks in position space. In Section~\ref{sec:Polyakov} below, we obtain some novel, explicit expressions for specific values  of $\Delta_\phi$ and $\Delta$. 

The dispersion relation is particularly well-suited to study perturbative theories. As mentioned above, the double discontinuity has double zeros at the dimensions of two-particles operators. This implies that their contribution to the double discontinuity  depends, at any given order in perturbation theory, entirely on lower order data. In this paper we consider two perturbative examples. As a warm-up exercise, we look at a  scalar field in AdS$_2$ and compute a four-point function of boundary operators,  previously bootstrapped using analytic functionals~\cite{Mazac:2018mdx}.   
The second setup that we consider is the defect CFT$_1$ living on the $1/2$ BPS supersymmetric Wilson line in planar $\mathcal{N}=4$ Super Yang Mills~\cite{Giombi:2017cqn}.
This defect CFT can be studied with integrability\cite{Kiryu:2018phb,Grabner:2020nis,Cavaglia:2021bnz,Cavaglia:2022qpg,Cavaglia:2023mmu} supersymmetric localization\cite{Giombi:2018qox,Giombi:2018hsx,Giombi:2020amn}, holography\cite{Giombi:2017cqn,Drukker:2005kx, Gomis:2006sb,Gomis:2006im}, and the conformal bootstrap \cite{Liendo:2016ymz,Liendo:2018ukf,Ferrero:2021bsb,Ferrero:2023gnu,Ferrero:2023znz,Bianchi:2021piu,Barrat:2021yvp,Barrat:2022psm,Bianchi:2022ppi}.
In particular, 
the authors of~\cite{Ferrero:2021bsb} managed to bootstrap the four-point function of the super-displacement multiplet~\cite{Liendo:2016ymz,Liendo:2018ukf} at fourth order in a strong t'Hooft coupling expansion, using  an Ansatz
 which involves polylogarithms and rational functions. A non-trivial step of their derivation is the resolution of operator mixing problem. 
Exploiting the analysis of~\cite{Ferrero:2021bsb}, in this paper we reproduce  their result using the dispersion relation. Bypassing the need of an Ansatz is the added value of our approach.  

In the future, it would be important to derive a general form of the dispersion kernel~\eqref{Kfinalin}, valid for any value of $\Delta_\phi$ and for non identical operators.
In order to do this, one would first need to correspondingly generalize the  Lorentzian inversion formula of~\cite{Mazac:2018qmi}.  
It is interesting to note that the double discontinuity of the dispersion kernel is a delta-function, see~\eqref{kerneldelta}, suggesting that the kernel is a combination of contact diagrams. One could then explore, at the level of the dispersion relation, possible implications of the observed symmetry of contact Witten diagrams~\cite{Rigatos:2022eos}.  Finally,  the dispersion relation~\eqref{dispersion} could well be used in other  perturbative setups, such as the defect CFT on the $1/2$ BPS Wilson line in ABJM~\cite{Aharony:2008ug,Drukker:2019bev,Drukker:2009hy,Bianchi:2017ozk,Bianchi:2020hsz} or line defects  in O($N$) models~\cite{Cuomo:2021kfm, Cuomo:2022xgw,Gimenez-Grau:2022czc,Bianchi:2022sbz,Gimenez-Grau:2022ebb,Bianchi:2023gkk,Dey:2024ilw}.

This paper proceeds as follows. In Section~\ref{sec:background} we sketch the background material necessary to formulate the problem. In Section~\ref{sec:kernel} we illustrate the derivation of the dispersion relation --  which is the main result of this paper -- and use it to compute new, explicit expressions for Polyakov blocks.  
In Section~\ref{sec:pert} we consider the perturbative setups of a scalar field in AdS$_2$ and the $1/2$ BPS supersymmetric Wilson line, reproducing known results 
with the use of the dispersion relation.

\section{CFT$_1$ kinematics and inversion formula}
\label{sec:background}

In this section we review some basic properties of correlators in a unitary CFT$_1$ and the Lorenzian inversion formula, presented in~\cite{Caron-Huot:2017vep} in the higher-dimensional case and extended to the one-dimensional case in~\cite{Simmons-Duffin:2017nub, Mazac:2018qmi}~\footnote{We refer the reader e.g. to Section 2 of~\cite{Mazac:2018qmi} for a thorough description of the features of unitary CFT$_1$s.}.
We  start by considering the four-point function of identical, scalar operators $\phi$ with dimension $\Delta_\phi$. The $SL(2, \mathbb{R})$ symmetry implies the structure
\begin{equation}\label{G}
   \langle \phi (x_1) \phi(x_2) \phi (x_3) \phi (x_4) \rangle = \frac{1}{(x_{12}\,x_{34})^{2\Delta_\phi}} \,\G(z) \,,
\end{equation}
where $z$ is the  
invariant cross-ratio
\be\label{chi}
 z=\frac{x_{12}\,x_{34}}{x_{13}\,x_{24}} \,,\qquad x_{ij}=x_i-x_j\,\,.
\ee
Considering the ordering  $x_1\!< x_2 \!< x_3\! < x_4\!\,$ on the line, conformal symmetry can be used to fix $x_1=0$, $x_3 = 1$, $x_4 = \infty$, so that $x_2 \equiv z \in (0, 1)$. 
Changing the ordering accordingly changes the range of $z$. Unlike the higher-dimensional case,  correlators obtained from~\eqref{G} via the exchange $1\leftrightarrow 2$ (and therefore defined in the regions $z<0$) and $2\leftrightarrow3$ (corresponding to $z>1$) are not related by analytic continuation. The correlator $\G(z)$ is piecewise analytic in the three disconnected regions $(-\infty,0)$, $(0,1)$ and $(1,\infty)$~\cite{Simmons-Duffin:2017nub,Mazac:2018qmi}, 
\begin{equation}\label{Gextended}
    \mathcal{G}(z) =
\begin{cases}
\mathcal{G}^{(-)}(z)\quad&\textrm{for }z\in(-\infty,0)\\
\mathcal{G}^{(0)}(z)\quad&\textrm{for }z\in(0,1)\\
\mathcal{G}^{(+)}(z)\quad&\textrm{for }z\in(1,\infty)\,.
\end{cases}
\end{equation}
As discussed in details in~\cite{Mazac:2018qmi}, in the case of identical operators Bose (or Fermi) symmetry can be used to show that the function $\G^{(0)}(z)$ above completely determines the correlator on the whole real line as
\begin{equation}\label{Gpm}
    \begin{split}
        &\mathcal{G}^{(-)}(z) = \mathcal{G}^{(0)}\Big(\textstyle \frac{z}{z-1}\Big) \, ,\\ 
         &\mathcal{G}^{(+)}(z) = \pm \, z^{2\Delta_\phi}\mathcal{G}^{(0)}\Big(\textstyle \frac{1}{z}\Big) \, , \\
    \end{split}
\end{equation}
where the plus sign is for bosons and the minus sign for fermions.
The exchange $x_1\leftrightarrow x_3$  (or equivalently, $x_2\leftrightarrow x_4$)  is the only true symmetry of the correlator, consistently mapping the interval $(0,1)$ to itself. It is expressed as the crossing relation 
\begin{equation}\label{crossing}
    \mathcal{G}^{(0)}(z) = \frac{z^{2\Delta_{\phi}}}{(1-z)^{2\Delta_{\phi}}}\mathcal{G}^{(0)}(1-z)~.
\end{equation}
The four-point function~\eqref{G} can be expanded using the OPE in two different channels\footnote{It is possible to show \cite{Hogervorst:2013sma} that the two conformal block expansions converge everywhere in the complex plane except on $(-\infty,0) \cup (1, \infty)$, thus providing an analytic continuation for $\mathcal{G}(z)$ to complex values of $z$.}
\begin{equation}\label{s-channel}
    \mathcal{G}^{(0)}(z) = \sum_{\Delta} a_{\Delta}  G_{\Delta} (z) = \frac{z^{2\Delta_{\phi}}}{(1-z)^{2\Delta_{\phi}}} \sum_{\Delta} a_{\Delta} G_{\Delta}(1-z) \, ,
\end{equation}
where the sum runs over $SL(2, \mathbb{R})$ primaries with dimension $\Delta$,  exchanged in the $\phi\times\phi$ OPE, $a_\Delta$ are the corresponding squared OPE coefficients and $G_\Delta (z)$ are the $SL(2, \mathbb{R})$ conformal blocks 
\begin{equation}
    G_\Delta (z) = z^{\Delta} {}_2F_1(\Delta,\Delta,2\Delta,z)\,.
\end{equation}
 Another useful decomposition is in terms of (Euclidean) conformal partial waves,  which  are regular eigenfunctions of the Casimir operator of $\mathfrak{sl}(2, \mathbb{R})$. 
 For $z\in (0,1)$, they can be written as a linear combination of conformal blocks and their shadow ($\Delta\to 1-\Delta$) blocks
\begin{equation}\label{CPW}
    \Psi_{\Delta}^{(0)} (z)= \kappa_{1-\Delta} G_{\Delta} (z) + \kappa_{\Delta} G_{1-\Delta} (z)\,,
    \qquad
    \kappa_{\Delta} = \frac{\sqrt{\pi} \,\Gamma(\Delta-\frac{1}{2})\Gamma(\frac{1-\Delta}{2})^2}{\Gamma (1-\Delta) \Gamma(\frac{\Delta}{2})^2}\,.
\end{equation}
In the equation above we used the same notation as in \eqref{Gextended}.
The representation theory of $SL(2,\mathbb{R})$~\cite{Harish-Chandra} 
shows that a complete and orthogonal set with respect to the invariant inner product $(f,g)=\int_{-\infty}^{\infty} dz \ z^{-2}\,f(z)\,g(z)$ is formed by 
 partial waves with (unphysical) complex dimensions $\Delta= \frac{1}{2} +i\alpha$ with $\alpha \in \mathbb{R}_+$,  referred to as the principal series representation,  and partial waves with discrete dimensions $\Delta=2m$, $m\in\mathbb{N}$. 
Explicitly, the orthogonality relations are 
\begin{equation}
    \begin{split}
    \big(\Psi_{\frac{1}{2} +i\alpha} (z)\,,\, \Psi_{\frac{1}{2} +i\beta} (z)\big)
      &= 2\pi \,n_{\frac{1}{2} +i\alpha}\delta(\alpha-\beta)\,, \qquad \alpha,\beta \in\mathbb{R}^+\\
     \big(\Psi_{2m} (z)\,,\, \Psi_{2n} (z)\big)  
     &= \frac{4\pi^2}{4m-1}\delta_{mn}\,,\qquad m,n \in \mathbb{N}
\end{split}
\end{equation} 
with $(\Psi_{1/2 +i\alpha} (z)\,,\, \Psi_m(z))=0$ and  $n_{\Delta} = 2 \kappa_{\Delta} \kappa_{1-\Delta}$.
The four-point function can be then decomposed as  
\begin{eqnarray}\label{gpartialwaves}
    \mathcal{G}(z) &=& \int_{\frac{1}{2}}^{\frac{1}{2}+i\infty} \frac{d \Delta }{2\pi i} \frac{I_{\Delta}}{n_{\Delta}} \Psi_{\Delta}(z) + 
    \sum_{m=0}^\infty \frac{4m-1}{4\pi^2} \tilde{I}_{2m} \Psi_{2m} (z) \\\label{G-CPW}
     &=& \int_{\frac{1}{2}-i\infty}^{\frac{1}{2}+i\infty} \frac{d\Delta}{2\pi i} \frac{I_{\Delta}}{2\kappa_{\Delta}} G_{\Delta}(z)  
   + \sum_{m=0}^\infty \frac{\Gamma^2(2m+2)}{2\pi^2\Gamma(4m+3)} \tilde{I}_{2m+2} G_{2m+2} (z) \, ,
\end{eqnarray}
where in the second line~\eqref{CPW} was used. 
The s-channel OPE decomposition~\eqref{s-channel} is recovered by closing the integration contour to the right, so that terms of the
OPE come from the  poles of the coefficients, or partial wave amplitude, defined as
\begin{equation}
    c(\Delta)=\frac{I_{\Delta}}{2\kappa_{\Delta}}\,,
\end{equation}
and from the terms of the discrete series. In particular the presence of an operator of dimension $\Delta$ in the OPE translates to a simple pole at $\Delta$ in $c(\Delta)$ with residue $-a_{\Delta}$. In general, $\tilde{I}_m$ is different from ${I}_{\Delta}|_{\Delta = m}$, unless there is no physical operator at $h=\Delta$. In this case they are equal, and the residue of the integral in~\eqref{G-CPW} coming from the zero of $\kappa_{\Delta}$ at $\Delta=m$ precisely cancels the corresponding term of the sum over the discrete series.
The coefficients $I_{\Delta}$ and $\tilde{I}_m$ may be obtained using orthogonality to invert equation~\eqref{gpartialwaves}~\footnote{The integrals~\eqref{inverseorthog} can be computed, in the regions $z<0$ and $z>1$, using~\cite{Mazac:2018qmi} 
\begin{equation}
    \begin{array}{ll}
\Psi_{\Delta}^{(-)}(z)=\Psi_{\Delta}^{(0)}\left(\frac{z}{z-1}\right) & \text { for } z \in(-\infty, 0) \\\label{inverseorthog}
\Psi_{\Delta}^{(+)}(z)=\frac{1}{2}\left[\Psi_{\Delta}^{(0)}(z+i \epsilon)+\Psi_{\Delta}^{(0)}(z-i \epsilon)\right] & \text { for } z \in(1, \infty)\,.
\end{array}
\end{equation}
} 
\begin{equation}\label{coeff-definition}
  \!\!\!\!\!\!      I_{\Delta}= \int_{-\infty}^{\infty} \!\!\!\!\!\! dw \ w^{-2} \mathcal{G}(w) \Psi_{\Delta}(w)\,,\qquad        
              \tilde{I}_{m}=\int_{-\infty}^{\infty} \!\!\!\!\!\! dw \ w^{-2} \mathcal{G}(w) \Psi_{m}(w)\,.
\end{equation}
However, in the case of identical operators, a far more powerful inversion  allows to reconstruct $\tilde{I}_m$~\cite{Simmons-Duffin:2017nub} and $I_\Delta$~\cite{Mazac:2018qmi} from the double discontinuity of the four-point function,
\begin{eqnarray}\label{LorInvDelta}
    I_\Delta &=& 2 \int_{0}^{1} dw\,w^{-2} H^{B/F}_{\Delta} (w) \, \dd[\mathcal{G}(w)] \, ,\\\label{LorInvm}
    \tilde{I}_m &=& \frac{4\Gamma^{2}(m)}{\Gamma(2m)}\int_{0}^{1} dw\,w^{-2} G_m (w)\, \dd[\mathcal{G}(w)] \, .
\end{eqnarray}
The double discontinuity is defined as 
\begin{equation}\label{ddisc}
    \dd[\mathcal{G}(z)] = \mathcal{G}(z)-\frac{\mathcal{G}^{\curvearrowleft}(z)+\mathcal{G}^{\curvearrowright
    }(z)}{2} \, ,
\end{equation}
where $\mathcal{G}^{\curvearrowleft}(z)$ is the value of $\mathcal{G}(z)$  moving counterclockwise around the branch cut at $z = 1$~\footnote{One can alternatively define the double discontinuity with respect to the other branch cut, at $z = 0$,  see for example~\cite{Paulos:2020zxx}, in which   case $\text{dDisc}[\mathcal{G}(z)]$  is  controlled by the s-channel  ($z\sim0$) OPE expansion. We follow here the convention of~\cite{Mazac:2018qmi}, and $\text{dDisc}[\mathcal{G}(z)]$ is controlled by the t-channel ($z\sim1$) OPE expansion, see \eqref{2.18}.} and viceversa for $\mathcal{G}^{\curvearrowright
    }(z)$.  
 For identical bosonic operators, \eqref{Gextended} implies 
\begin{equation}
\mathcal{G}^{\curvearrowleft}(z)=\mathcal{G}^{(+)}(z+i \epsilon)\,,\qquad\qquad\qquad
    \mathcal{G}^{\curvearrowright}(z)=\mathcal{G}^{(+)}(z-i \epsilon) \, .
\end{equation}
Evaluating the double discontinuity of the correlator commutes with its OPE expansion in the t-channel~\cite{Mazac:2018qmi}, so that
\begin{equation}\label{2.18}
    \dd \big[\mathcal{G}(z)]= \sum_\Delta a_\Delta \dd \big[\textstyle\frac{z^{2\Delta_{\phi}}}{(1-z)^{2\Delta_{\phi}}} G_{\Delta}(1-z)\big] \, .
\end{equation}
For a bosonic correlator the contribution of a single conformal block reads 
\begin{equation}\label{ddiscblockbos}
        \dd \big[\textstyle\frac{z^{2\Delta_{\phi}}}{(1-z)^{2\Delta_{\phi}}} G_{\Delta}(1-z)\big] = 2\sin^{2}{\frac{\pi}{2} (\Delta-2\Delta_{\phi})}\frac{z^{2\Delta_{\phi}}}{(1-z)^{2\Delta_{\phi}}} G_{\Delta}(1-z) ~,
\end{equation}
while for a fermionic one
\begin{equation}\label{ddiscblockfer}
        \dd \big[\textstyle\frac{z^{2\Delta_{\phi}}}{(1-z)^{2\Delta_{\phi}}} G_{\Delta}(1-z)\big] = 2\cos^{2}{\frac{\pi}{2} (\Delta-2\Delta_{\phi})}\frac{z^{2\Delta_{\phi}}}{(1-z)^{2\Delta_{\phi}}} G_{\Delta}(1-z) ~.
\end{equation}
Notice that the double discontinuity of conformal blocks, and of their derivatives with  respect to $\Delta$, vanishes for the case of two-particle operators, which have dimensions $\Delta=2\Delta_\phi +2n$ in the bosonic case and $\Delta=2\Delta_\phi +2n+1$ in the fermionic one.\\
The functions $H^{B/F}_{\Delta} (w)$ in \eqref{LorInvDelta} are inversion kernels, respectively for the bosonic and fermionic case, that can be determined requiring consistency between the inversion~\eqref{LorInvDelta} and the definition~\eqref{coeff-definition},  as discussed thoroughly in~\cite{Mazac:2018qmi}. In particular, together with  being holomorphic in $w\notin (1,\infty)$ -- which implies the absence of poles in $w=0$ -- they should satisfy the constraints $H^{B/F}_{\Delta} (w)=H^{B/F}_{\Delta} (w/(w-1))$ and
\begin{equation}\label{Hequation}
\begin{split}
    &z^{2\Delta_\phi-2}H^{B/F}_\Delta(z)+(1-z)^{2\Delta_\phi-2}H^{B/F}_\Delta(1-z) \pm\frac{H^{B/F}_\Delta\left(\frac{1}{z}+i\epsilon\right)+H^{B/F}_\Delta\left(\frac{1}{z}-i\epsilon\right)}{2} =\\
=&z^{2\Delta_\phi-2}\Psi_\Delta^{(0)}(z)+(1-z)^{2\Delta_\phi-2}\Psi_\Delta^{(0)}(1-z) \pm\frac{\Psi_\Delta^{(0)}\left(\frac{1}{z}+i\epsilon\right)+\Psi_\Delta^{(0)}\left(\frac{1}{z}-i\epsilon\right)}{2}\,.
\end{split}
\end{equation}
The  explicit form of such  inversion kernels is  only known in the case of identical bosons (fermions) with integer (half integer) conformal dimension~\cite{Mazac:2018qmi}, and reads~\footnote{See also the general expression for the Taylor expansion of $H_{\Delta}^{B/F} (z)$ around $z=0$, formula (4.8) in~ \cite{Mazac:2018qmi}.} 
\begin{eqnarray}\label{Hexpl}
     H_{\Delta}^{B/F} (w) &=& \pm\frac{2 \pi}{ \sin (\pi \Delta)} \left[w^{2-2\Delta_{\phi}} p_{\Delta}(w)+\big(\textstyle\frac{w}{w-1}\big)^{2-2\Delta_{\phi}} p_{\Delta}(\textstyle\frac{w}{w-1}\big)+q^{\Delta_{\phi}}_{\Delta}(w))\right] \, , \\\label{pdef}
     p_{\Delta}(w) &=&{}_2F_1(\Delta,1-\Delta,1,w) \, ,  \\\label{qdef}
    q^{\Delta_{\phi}}_{\Delta}(w) &=& a^{\Delta_{\phi}}_{\Delta}(w) +b^{\Delta_{\phi}}_{\Delta}(w) \log(1-w)\,.
\end{eqnarray}
In~\eqref{qdef},   $a^{\Delta_{\phi}}_{\Delta}(w)$ and $b^{\Delta_{\phi}}_{\Delta}(w)$ are  polynomials in $\Delta$\footnote{Notice that \eqref{CPW} and \eqref{coeff-definition} imply $H_{\Delta}^{B/F} (w)=H_{1-\Delta}^{B/F} (w)$, therefore $a^{\Delta_{\phi}}_{\Delta}(w)$ and $b^{\Delta_{\phi}}_{\Delta}(w)$ are actually polynomials in $\Delta(\Delta-1)$.} and $w$,
\begin{equation}\label{qcoeff}
    \begin{split}
        &a^{\Delta_{\phi}}_{\Delta}(w)= \sum_{m=0}^{2\Delta_\phi-2} \sum_{n=0}^{2\Delta_\phi-4} \alpha_{m,n} \, w^{m+2-2\Delta_{\phi}} \Delta^n (\Delta-1)^n \, ,\\
        &b^{\Delta_{\phi}}_{\Delta}(w)= \sum_{m=0}^{2\Delta_\phi-2} \sum_{n=0}^{2\Delta_\phi-4} \beta_{m,n} \, w^{m+2-2\Delta_{\phi}} \Delta^n (\Delta-1)^n \, .
    \end{split}
\end{equation}
The coefficients $\alpha_{m,n}$ and $\beta_{m,n}$ above must be determined, for each given $\Delta_{\phi}$, from the requirement that $ H^{B/F}_{\Delta} (w)$ has no pole at $w=0$. 
The first few examples read~\cite{Mazac:2018qmi}
\begin{equation}
    \begin{split}
        &\quad a^{1}_{\Delta}(w) = 0\,,\quad b^{1}_{\Delta}(w) = 0 \, ,\\
        &\quad a^{2}_{\Delta}(w) = w^2+2 w-2\,,\quad b^{2}_{\Delta}(w) = 0 \, , \\
        &\quad a^{1/2}_{\Delta}(w) = 0\,,\quad b^{1/2}_{\Delta}(w) = 0 \, ,\\
        &\quad a^{3/2}_{\Delta}(w) =\left(2 \Delta ^2-2 \Delta -1\right) w\,,\quad b^{3/2}_{\Delta}(w) = 0 \, .
    \end{split}
\end{equation}
It is worth noticing that, while in higher dimensions the inversion kernel  does not depend on the external dimensions $\Delta_\phi$ and it is simply a conformal block~\cite{Caron-Huot:2017vep}, in $d=1$ the dependence of  $H_{\Delta}^{B/F} (w)$ on  $\Delta_\phi$ is non-trivial. On the other hand, in the present case one positive feature is that the inversion kernel gives rise to a correlator which is manifestly crossing symmetric. This can be seen from  inverting a single conformal block in the t-channel, which returns the coefficient  of a fully crossing-symmetric sum of exchange Witten  diagrams in AdS$_2$, including the direct s-channel exchange, as discussed  in~\cite{Mazac:2018qmi}~\footnote{See Section 5 there.}.

 In unitary CFTs, four-point functions are bounded in the Regge limit~\cite{Maldacena:2015waa,Caron-Huot:2017vep}, i.e. one has~\cite{Mazac:2018ycv} 
\begin{equation}\label{bounded}
  \Big(\frac{1}{2}+i t\Big)^{-2\Delta_{\phi}}\,\mathcal{G}\Big(\frac{1}{2}+i t\Big)<\infty \ \ \ \text{for } t\rightarrow \infty \, .
\end{equation}
With the kernels~\eqref{Hexpl}, both the inversion formulas~\eqref{LorInvDelta}-\eqref{LorInvm}  hold, in the fermionic case, for any Regge-bounded four-point function. In the bosonic case, the  inversion~\eqref{LorInvDelta} holds only for functions which are Regge super-bounded,  for which namely it holds
\begin{equation}\label{superbounded}
 \left(\frac{1}{2}+i t\right)^{-2\Delta_{\phi}}\,\mathcal{G}\left(\frac{1}{2}+i t\right) \sim t^{-1-\epsilon} \ \ \ \text{for } t\rightarrow \infty \, ,
\end{equation}
with $\epsilon>0$.
In general, the behaviour of the correlator in the Regge limit is related to the behaviour at $w=0$ of the inversion kernel \cite{Mazac:2018qmi},
\begin{equation}\label{ReggeH}
     \left(\frac{1}{2}+i t\right)^{-2\Delta_{\phi}}\,\mathcal{G}\left(\frac{1}{2}+i t\right) \sim t^{n} \ \text{for } t\rightarrow \infty  \ \Longrightarrow \ H_\Delta(w) \sim w^{2+2n} \ \text{for } w\rightarrow0 \, .
\end{equation}
Indeed $H^B _\Delta (w) \sim w^0$ and $H^F _\Delta (w)\sim w^2$. In order to obtain an inversion formula valid also for Regge-bounded bosonic correlators, one has to improve the behaviour at $w=0$ of the inversion kernel $H^B _\Delta (w)$ by subtracting any function that satisfies all the properties of $H^B _\Delta (w)$,  as well as~\eqref{Hequation} with vanishing right-hand side. There are two infinite families of such functions, $\widehat{H}_{n, 2}^{B}(w)$ and $\widehat{H}_{n, 1}^{B}(w)$, which can be obtained by expanding $H^B _\Delta (w)$ near $\Delta = 2 \Delta_\phi+2n$
\begin{equation}\label{Hseries}
    \frac{H_{\Delta}^{B}(w)}{\kappa_\Delta} = \frac{\widehat{H}_{n, 2}^{B}(w)}{\left(\Delta-2\Delta_\phi-2n \right)^2}+\frac{\widehat{H}_{n, 1}^{B}(w)}{\Delta-2\Delta_\phi-2n}+O(1)\,,\qquad \ \ \ \ \ \Delta \rightarrow 2\Delta_\phi+2n \, .
\end{equation} 
Using for instance $\widehat{H}_{0,2}^{B}(w)$, one obtains for bounded bosons \cite{Mazac:2018qmi}
\begin{equation}\label{Hboson}
    H^{\text{bd}}_{\Delta}(w) \equiv H_{\Delta}^{B}(w)-\frac{\pi^2 2^{2\left(\Delta_\phi-1\right)} \Gamma\left(\Delta_\phi+\frac{1}{2}\right)}{\Gamma\left(\Delta_\phi\right)^3 \Gamma\left(2 \Delta_\phi-\frac{1}{2}\right)} \frac{\Gamma\left(\Delta_\phi-\frac{\Delta}{2}\right)^2 \Gamma\left(\Delta_\phi-\frac{1-\Delta}{2}\right)^2}{\Gamma\left(1-\frac{\Delta}{2}\right)^2 \Gamma\left(1-\frac{1-\Delta}{2}\right)^2} \frac{2 \pi}{\sin (\pi \Delta)} \widehat{H}_{0,2}^{B}(w) \, ,
\end{equation}
We stress that the coefficient in front of $\widehat{H}_{0,2}^{B}(w)$ is fixed by demanding that $ H^{\text{bd}}_{\Delta}(w)  \sim w^2$ for $w \rightarrow 0$.
It turns out that $\widehat{H}_{0,2}^{B}(w)$ has a pole in $w=1$ for all $\Delta_\phi$, which may spoil the convergence of the integral \eqref{LorInvDelta}. Therefore, in addition to the kernel redefinition \eqref{Hboson}, one may have to define a regularized correlator $\mathcal{G}^{\text{reg}}(z)$ by subtracting  a crossing symmetric and Regge-bounded function from $\mathcal{G}(z)$, with the aim of getting rid of the singularity in the integral. In general, one defines the regularized correlator
\begin{equation}\label{defreg}
    \mathcal{G}^{\text{reg}}(z) = \mathcal{G}(z) - \sum_{\Delta < 2\Delta_\phi} a_\Delta \mathcal{P}^{\Delta_\phi}_{\Delta}(z) \, ,
\end{equation}
where $\mathcal{P}^{\Delta_\phi}_{\Delta}(z)$ are Polyakov blocks, which are crossing symmetric and Regge-bounded functions with the same double discontinuity~\eqref{ddiscblockbos},\eqref{ddiscblockfer} as the conformal blocks, see below Section~\ref{sec:Polyakov}. For instance, in the bosonic case
\begin{equation}
        \dd \big[\mathcal{P}^{\Delta_\phi}_{\Delta}(z)\big] =  2 \sin^{2}{\frac{\pi}{2} (\Delta-2\Delta_{\phi})}\frac{z^{2\Delta_{\phi}}}{(1-z)^{2\Delta_{\phi}}} G_{\Delta}(1-z) \, .
\end{equation}
Therefore~\footnote{In some sense, the necessity of a subtraction means that the inversion formula misses the contribution from some low-dimensional operators. Morally, this is then the one-dimensional equivalent of the low-spin ambiguities which affect the higher-dimensional case~\cite{Caron-Huot:2017vep}.}
\begin{equation}
   \begin{split}
        \dd \big[\mathcal{G}^{\text{reg}}(z)\big] &= \sum_{\Delta > 2\Delta_\phi} 2 a_\Delta \sin^{2}{\frac{\pi}{2} (\Delta-2\Delta_{\phi})}\frac{z^{2\Delta_{\phi}}}{(1-z)^{2\Delta_{\phi}}} G_{\Delta}(1-z) \\
    &\sim (1-z)^\epsilon \quad \text{with } \epsilon >0 \, ,
   \end{split}
\end{equation}
and the integral in \eqref{LorInvDelta} converges.  While the  definition~\eqref{defreg} of $\mathcal{G}^{\text{reg}}(z)$ works in full generality, it is  hard to use,  because Polyakov blocks are in general complicated functions. Therefore, in concrete applications it is often more convenient to use different, ad hoc subtractions. We shall see examples of these in Section \ref{sec:pert}.
All in all, the inversion formula for bounded bosonic four-point functions reads \cite{Mazac:2018qmi}
\begin{equation}
I_{\Delta}=  2 \int_0^1 \frac{dw}{w^2} H^{\text{bd}}_{\Delta}(w) \text{dDisc}\mathcal{G}^{\text{reg}}(w)
 +\lim _{\rho \rightarrow 0} \int_{C_\rho^{+}} \frac{dw}{w^2}  H^{\text{bd}}_{\Delta}(w) \ \mathcal{G}^{\text{reg}}(w)+\lim _{\rho \rightarrow 0} \int_{C_\rho^{-}} \frac{dw}{w^2}  H^{\text{bd}}_{\Delta}(w) \mathcal{G}^{\text{reg}}(w) \, ,
\end{equation}
where $C_\rho^{\pm}$ are semicircular contours of radius $\rho$ centered in $w=1$ going above and
below the real axis, which we have to introduce to avoid the pole generated by $\widehat{H}_{0,2}^{B}(w)$.
The strategy used to obtain the Regge-bounded inversion kernel can be obviously generalized if one needs, as we do in this paper, to a kernel for correlators \emph{unbounded} in the Regge limit, such as those that usually appear in perturbation theory~\footnote{The bound on the growth of the correlator in the Regge limit is a non perturbative statement that can fail order by order in perturbation theory.}.  Given a certain behaviour in the Regge limit, one can obtain the corresponding inversion kernel by defining
\begin{equation}\label{Hunbounded}
   \!\!\! H^{\text{unbd}}_{\Delta}(w) \equiv \textstyle H_{\Delta}^{B}(w)- \sum\limits_{m,n} A_{m,n} \, \widehat{H}_{m,2}^{B}(w) \frac{2 \pi\,\Delta^n (\Delta-1)^n}{\sin (\pi \Delta)} - \sum\limits_{m,n} B_{m,n} \, \widehat{H}_{m,1}^{B}(w) \frac{2 \pi\,\Delta^n (\Delta-1)^n}{\sin (\pi \Delta)}  
\end{equation}
and fixing the unknown coefficients $ A_{m,n},  B_{m,n}$ by imposing the desired behaviour at small $w$ as in \eqref{ReggeH}. The more $\mathcal{G}(z)$ diverges in the Regge limit, the more subtractions will be needed and the stronger the singularities at $w=1$ will be\footnote{The functions $\widehat{H}_{n,2}^{B}(w)$ have poles at $w=1$ of increasing order as $n$ increases, while $\widehat{H}_{n,1}^{B}(w)$ have also logarithmic singularities.}. For example, for a correlator that diverges linearly in the Regge limit two subtractions are necessary. We shall see related examples in Section~\ref{sec:pert}. 

\section{The dispersion relation}
\label{sec:kernel}

We can now discuss the derivation of a dispersion relation for one-dimensional CFTs from the Lorentzian inversion formulae \eqref{LorInvm} \eqref{LorInvDelta}. 
A dispersion relation for CFT$_1$s already appeared in~\cite{Paulos:2020zxx}, where it was constructed from so-called master functionals, which are analytic extremal functionals acting on the crossing equation~\footnote{See  \cite{Mazac:2016qev,Mazac:2018mdx,Mazac:2018ycv} for related work on analytic functionals in one-dimensional CFTs}. The main result of \cite{Paulos:2020zxx} is a set of functional equations, or alternatively a single Fredholm integral equation of the second kind~\footnote{Notice that a different convention is used in~\cite{Paulos:2020zxx}, where $\mathcal{G}^{\text{there}}(w)\equiv z^{-2\Delta_\phi} \mathcal{G}(z)^{\text{there}}$. Also, the double discontinuity is evaluated around $z = 0$ rather than around $z = 1$ as we do here.}, 
which implicitly determine  the kernel $K_{\Delta_\phi}(z,w)$. These equations can only be solved numerically for general $\Delta_\phi$. For integer or half-integer dimensions, the author of~\cite{Paulos:2020zxx} found an Ansatz for a solution in terms of certain rational functions and logarithms, which must be fixed case by case imposing a series of conditions. We refer to Appendix A of \cite{Paulos:2020zxx} for more details on this approach. Here we provide an alternative derivation of the dispersion relation in the case of bosonic (fermionic) operators with integer (half-integer) dimensions, which turns out to be simpler 
and allows us to obtain a more explicit formula for all integer or half-integer $\Delta_\phi$. We follow the strategy that worked in the higher-dimensional case~\cite{Carmi:2019cub}. 
 
\subsection{Explicit form of the dispersion relation} 
\label{sec:kernelexpl}
 
We start by considering  the case of Regge-(super)bounded (bosonic) fermionic correlators.
Starting from the expansion~\eqref{G-CPW}, we plug in the inversion formulas~\eqref{LorInvDelta}-\eqref{LorInvm} and exchange the order of integration
\begin{equation}
\begin{split}
    \mathcal{G}(z) &=     \int_{0}^{1} dw \ w^{-2} \text{dDisc}[\mathcal{G}(w)]   \int_{\frac{1}{2}-i\infty}^{\frac{1}{2}+i\infty} \frac{d \Delta}{2\pi i} \frac{ H^{B/F}_{\Delta}(w) }{\kappa_{\Delta}} \, G_{\Delta}(z)  \\ &+
    \int_{0}^{1} dw \ w^{-2} \text{dDisc}[\mathcal{G}(w)]  \ \sum_{m =0}^{\infty} \frac{2 \Gamma (2m+2)^4}{\pi ^2 \Gamma (4m+4) \Gamma (4m+3)} G_{2m+2}(w)  G_{2m+2} (z)  \\ \label{kerneldef}
    &\equiv \int_{0}^{1} dw \ w^{-2} \text{dDisc}[\mathcal{G}(w)] K_{\Delta_\phi}(z,w)\,,
\end{split}    
\end{equation}
the aim being to work out the kernel $K_{\Delta_\phi}(z,w)$. As mentioned above,  the latter is crossing symmetric and explicitly depends on the dimension $\Delta_\phi$ of the external operators. It also inherits from $\mathcal{G}(z)$ its Regge-boundedness. Notice that applying $\dd$ to both sides of~\eqref{kerneldef} one obtains
\begin{equation}\label{kerneldelta}
    w^{-2}\text{dDisc}[K_{\Delta_\phi}(z,w)]  
    = \delta(z-w)\,.
\end{equation}
Using~\eqref{Hexpl}, $K_{\Delta_\phi}(z,w)$ is  explicitly defined by
\begin{align}\nonumber
        K_{\Delta_\phi}(z,w) &= \pm \int_{\frac{1}{2}-i\infty}^{\frac{1}{2}+i\infty} \frac{d \Delta}{2\pi i} \frac{G_{\Delta}(z)}{\kappa_{\Delta}}  \frac{2 \pi}{ \sin (\pi \Delta)} \Big[w^{2-2\Delta_{\phi}} p_{\Delta}(w)+\textstyle(\frac{w}{w-1})^{2-2\Delta_{\phi}} p_{\Delta}(\textstyle\frac{w}{w-1})+q^{\Delta_{\phi}}_{\Delta}(w)\Big]  \\\nonumber
        &\quad+ \sum_{m=0}^\infty \frac{2 \Gamma (2m+2)^4}{\pi ^2 \Gamma (4m+4) \Gamma (4m+3)} G_{2m+2}(w)  G_{2m+2} (z)
         \\\label{Ksplit}
        &\equiv  K_{\Delta_\phi}^{p}(z,w)+K_{\Delta_\phi}^q (z,w) +K^{\text{discrete}}(z,w)\, ,
\end{align}
where the three contributing terms in the last line are defined below.
The discrete contribution does not depend on the external dimension $\Delta_\phi$, it is identical for both fermions and bosons and it reads
\begin{equation}
    \begin{split}
         K^{\text{discrete}}(z,w)&\equiv \sum_{m =0}^{\infty} \frac{2 \Gamma (2m+2)^4}{\pi ^2 \Gamma (4m+4) \Gamma (4m+3)} G_{2m+2}(w)  G_{2m+2} (z)  \\\label{Kdiscrete}
        &= \sum_{m =0}^{\infty} \frac{8(4m+3)}{\pi^{2}} Q_{2m+1}\textstyle(\frac{2}{w}-1)Q_{2m+1}(\frac{2}{z}-1) \, ,
    \end{split}
\end{equation}
where $Q_{n}(z)$ are Legendre functions of the second kind and we used that
\begin{equation}\label{blocklegendre}
    G_{2m+2} = \frac{2^{4 m+4} \Gamma (\textstyle2 m+\frac{1}{2}+2) }{\sqrt{\pi } \Gamma (2 m+2)} Q_{2 m+1}\textstyle(\frac{2}{z}-1) \, .
\end{equation}
The discrete sum can be then evaluated using 
\begin{equation}\label{Qintrep1}
    Q_{2 m+1}\textstyle(\frac{2}{z}-1) = \int_{-1}^{1} dv \ \frac{P_{2 m+1}(v)}{2 (-v+\frac{2}{z}-1)} \, ,
\end{equation}
and 
\begin{equation}\label{sumP}
    \sum_{m=0}^{\infty}(4 m+3)  P_{2 m+1}(x) P_{2 m+1}(y) = \delta (x-y)-\delta (x+y) \, ,
\end{equation}
where $P_n (z)$ are Legendre functions of the first kind. We find
\begin{align}\nonumber
        K^{\text{discrete}}(z,w)&=\sum_{m =0}^{\infty} \frac{8(4m+3)}{\pi^{2}} Q_{2m+1}\textstyle(\frac{2}{w}-1)Q_{2m+1} (\frac{2}{z}-1)\\\nonumber
        &=\textstyle\int_{-1}^{1} dv \int_{-1}^{1} du \, \frac{1}{2(\frac{2}{w}-1-v)}    \frac{1}{2(\frac{2}{z}-1-u)} \sum_{m=0}^{\infty} (4m+3) P_{2m+1}(v)P_{2m+1}(u)\\
        &=\frac{w\, z^2 (w-2) \log (1-w)}{\pi ^2 (w-z) (w+z-wz)}-\frac{z\,w^2 (z-2)  \log (1-z)}{\pi ^2 (w-z) (w+z-wz)} \, .
\end{align}
The second term in \eqref{Ksplit} is defined by
\begin{equation}\label{Kpdef}
   K_{\Delta_\phi}^p (z,w) \equiv \pm \int_{\frac{1}{2}-i\infty}^{\frac{1}{2}+i\infty} \frac{d \Delta}{2\pi i} \frac{ G_{\Delta}(z)}{\kappa_{\Delta}} \frac{2 \pi}{ \sin (\pi \Delta)}\textstyle \Big[w^{2-2\Delta_{\phi}} p_{\Delta}(w)+ (\frac{w}{w-1})^{2-2\Delta_{\phi}} p_{\Delta}(\frac{w}{w-1})\Big] \, , 
\end{equation}
with $p_{\Delta}(z)$  in \eqref{pdef}.
Closing the contour on the right and using the residue theorem, one obtains
\begin{eqnarray}\nonumber
&&\!\!\!\!\!\!\!\!\!\! \!\!\!\!\! K_{\Delta_\phi}^p (z,w) = \textstyle \mp \sum_{m =0}^{\infty}  \partial_m \big [\frac{2 (4 m+3)}{\pi ^2}  Q_{2 m+1} (\frac{2}{z}-1 )  \big(\,w^{2-2 \Delta \phi } P_{2 m+1}(1-2 w)+(\textstyle w \!\rightarrow\! \frac{w}{w-1} ) \,\big)\big ] \\\textstyle        \label{Kpcomp}
&&\!\!\!\!\!\!\!\!\!\!\!\!\!\!\!= \pm \frac{z^2 }{\pi ^2}  \Big[\,\textstyle\!\!\log(1\!-\!w)\frac{(1-2w) w^{2-2 \Delta \phi }}{(w-1) w z^2+z-1}    + \frac{\log (1-z) }{z} \textstyle \frac{w^{2-2 \Delta \phi }}{w z-1}  + \log (z)  \frac{(1-2w) w^{2-2 \Delta \phi }}{(w-1) w z^2+z-1}+(\textstyle w\! \rightarrow \!\frac{w}{w-1} )\,\Big] \, ,\end{eqnarray}
where the integral representations of Legendre functions 
\begin{equation}
    \begin{split}
         P_n(z)&=\frac{2^n}{\pi }  \int_{-\infty }^{\infty } du \frac{(z+i u)^n}{\left(u^2+1\right)^{n+1}}  \\
          Q_{\nu }^{\mu }(z)&=\frac{1}{\Gamma (\nu +1)}e^{\pi  i \mu } 2^{-\nu -1} \left(z^2-1\right)^{\mu /2} \Gamma (\mu +\nu +1) \int_{-1}^1  dt\left(1-t^2\right)^{\nu } (z-t)^{-\mu -\nu -1}   
           \end{split}
\end{equation}
were used to perform the sum in the first line of~\eqref{Kpcomp}, by exchanging the order of sum and integral.  \\
The last contribution to the kernel $K_{\Delta_\phi}(z,w)$ in \eqref{Ksplit}  is
\begin{equation}
     K_{\Delta_\phi}^q (z,w)\equiv \int_{\frac{1}{2}-i\infty}^{\frac{1}{2}+i\infty} \frac{d \Delta}{2\pi i} \frac{ G_{\Delta}(z)}{\kappa_{\Delta}} \frac{2 \pi}{ \sin (\pi \Delta)} \, q^{\Delta_{\phi}}_{\Delta}(w) \, ,
\end{equation}
where $q^{\Delta_{\phi}}_{\Delta}(w)$ is defined in \eqref{qdef}.
Using the explicit structure of  $q^{\Delta_{\phi}}_{\Delta}(w)$, see \eqref{qdef} and \eqref{qcoeff}, we find that
\begin{equation}\label{Kq}
   \textstyle K_{\Delta_\phi}^q (z,w) =  \sum\limits_{m=0}^{2\Delta_\phi - 2} \sum\limits_{n=0}^{2\Delta_\phi - 4}  (\alpha_{m,n}+\beta_{m,n}\log(1-w)) w^{m+2-2\Delta_\phi} \int_{\frac{1}{2}-i\infty}^{\frac{1}{2}+i\infty} \frac{d \Delta}{2\pi i} \frac{ G_{\Delta}(z)}{ \kappa_{\Delta}} \frac{2 \pi }{ \sin (\pi \Delta)} \Delta^n (\Delta-1)^n \, ,
\end{equation}
where   we stress that $\alpha_{m,n}$ and $\beta_{m,n}$ are a finite number of coefficients, fixed case by case by demanding that $ H^{B/F}_{\Delta}(w) $ has no poles in $w=0$. 
The integral in \eqref{Kq}  can be now trivially evaluated considering the action of the Casimir operator $\mathcal{C} =z^2 (1-z) \partial^2 -z^2 \partial$ under the integral over $\Delta$ in the definition~\eqref{Kpdef} of  $K_{\Delta_\phi =1}^p(z,w=0)$, and using  $\mathcal{C} G_\Delta(z) = \Delta (\Delta-1) G_\Delta(z)$.  
Namely, one obtains
\begin{equation}\label{casimir}        
\textstyle\int_{\frac{1}{2}-i\infty}^{\frac{1}{2}+i\infty} \frac{d \Delta}{2\pi i} \frac{G_{\Delta}(z)}{\kappa_{\Delta}} \frac{2 \pi }{ \sin (\pi \Delta)} \Delta^n (\Delta-1)^n =\mathcal{C}^n \left[K_{\Delta_\phi =1}^p (z,0)\right] =       \textstyle \mathcal{C}^n \left[-\frac{2}{\pi^2} \left(\frac{z^2 \log (z)}{1-z}+z \log (1-z)\right)\right]\,.
\end{equation}
All in all, the kernel reads 
\begin{align}\label{Kfinal}
        K_{\Delta_\phi}(z,w) &= \frac{w\, z^2 (w-2) \log (1-w)}{\pi ^2 (w-z) (w+z-wz)}-\frac{z\,w^2 (z-2)  \log (1-z)}{\pi ^2 (w-z) (w+z-wz)}\\\nonumber
        &\pm \frac{z^2 }{\pi ^2}  \Big[\,\textstyle\!\!\log(1\!-\!w)\frac{(1-2w) w^{2-2 \Delta_\phi }}{(w-1) w z^2+z-1}    + \frac{\log (1-z) }{z} \textstyle \frac{w^{2-2 \Delta_\phi }}{w z-1}  + \log (z)  \frac{(1-2w) w^{2-2 \Delta_\phi }}{(w-1) w z^2+z-1}+(\textstyle w\! \rightarrow \!\frac{w}{w-1} )\,\Big]\\\nonumber
        & \textstyle +\sum\limits_{m=0}^{2\Delta_\phi - 2} \sum\limits_{n=0}^{2\Delta_\phi - 4}  (\alpha_{m,n}+\beta_{m,n}\log(1-w)) w^{m+2-2\Delta_\phi} \,\mathcal{C}^n \left[\frac{2}{\pi^2} \left(\frac{z^2 \log (z)}{1-z}+z \log (1-z)\right)\right] \, .
\end{align}
As discussed, the kernel inherits crossing symmetry from the inversion formula \cite{Mazac:2018qmi}, so that
\be
 K_{\Delta_\phi}(z,w) = \frac{z^{2\Delta_\phi}}{(1-z)^{2\Delta\phi}}\, K_{\Delta_\phi}(1-z,w)\,.
\ee
 This can be usefully rewritten as
\begin{eqnarray}\nonumber
         &&\!\!\!\!\!\textstyle\sum\limits_{m=0}^{2\Delta_\phi - 2} \sum\limits_{n=0}^{2\Delta_\phi - 4} (\alpha_{m,n}+\beta_{m,n}\log(1-w)) w^{m+2-2\Delta_\phi} \Big(\mathcal{C}^n \Big[\frac{2}{\pi^2} \Big(\frac{z^2 \log (z)}{1-z}+z \log (1-z)\Big)\Big] \!\!-\text{crossing} \Big) = \\\label{Kcrossing}
        & &= \frac{z^{2\Delta_\phi}}{(1-z)^{2\Delta_\phi}} \left( K^{\text{discrete}}(1-z,w) + K_{\Delta_\phi}^{p}(1-z,w) \right) -K^{\text{discrete}}(z,w) - K_{\Delta_\phi}^{p}(z,w) \, ,
\end{eqnarray}
 a system of equations which can be used to determine $\alpha_{m,n}$ and $\beta_{m,n}$.   One could of course still find the latter as discussed below~\eqref{qcoeff}, and the equation above will be for them trivially satisfied.  Clearly, once all the coefficients are fixed, the kernel   $K_{\Delta_\phi}(z,w)$  satisfies crossing symmetry~\eqref{crossing} and boundedness~\eqref{bounded} in $z$, and its double discontinuity is a delta-function as in~\eqref{kerneldelta}. 
 Plotting \eqref{Kfinal} for several values of $\Delta_\phi$ and $(z,w)\in (0,1)$ we found  that the kernel is always positive. 
 In particular, setting $\Delta_\phi=1/2$ we reproduce the explicit result in Appendix A of \cite{Paulos:2020zxx}, modulo different conventions.

The   dispersion kernel ~\eqref{Kfinal} is valid only for super-bounded bosonic correlators~\eqref{superbounded}. 
To extend the result to bounded bosonic correlators, one should use in~\eqref{kerneldef} the improved inversion kernel~\eqref{Hboson}. The dispersion relation then reads
\begin{equation}\label{dispersionboson}
 \begin{split}
        \mathcal{G}^{\text{reg}} (z) &=  \int_{0}^{1} \frac{dw}{w^2} K^{\text{bd}}_{\Delta_\phi}(z,w) \, \text{dDisc}\mathcal{G}^{\text{reg}}(w) + \\
& +\lim _{\rho \rightarrow 0} \int_{C_\rho^{+}} \frac{dw}{2w^2}  K^{\text{bd}}_{\Delta_\phi}(z,w) \mathcal{G}^{\text{reg}}(w)+\lim _{\rho \rightarrow 0} \int_{C_\rho^{-}} \frac{dw}{2w^2} K^{\text{bd}}_{\Delta_\phi}(z,w) \mathcal{G}^{\text{reg}}(w) \, .
 \end{split}
\end{equation}
with
\begin{align}\label{Kimpr}
     &K^{\text{bd}}_{\Delta_\phi}(z,w) =K_{\Delta_\phi}(z,w)-\widehat{H}_{0,2}^{B}(w) \, \sum\limits_{n=0}^{2\Delta_\phi -2} A_{n} \,\mathcal{C}^n \left[\frac{2}{\pi^2} \left(\frac{z^2 \log (z)}{1-z}+z \log (1-z)\right)\right]\, .
\end{align}
For the  extra term we have used  the same strategy  as for $K^q_{\Delta_\phi}$. One exploits the fact that for every integer $\Delta_\phi$ the ratio of gamma functions appearing in \eqref{Hboson} reduces to a polynomial in $\Delta (\Delta -1)$, namely
\begin{align}\label{boundcoeff}
    \frac{\pi^2 2^{2(\Delta_\phi-1)} \Gamma (\Delta_\phi+\frac{1}{2}  )}{\Gamma (\Delta_\phi  )^3 \Gamma (2 \Delta_\phi-\frac{1}{2}  )} 
    \textstyle\frac{\Gamma (\Delta_\phi-\frac{\Delta}{2}  )^2 \Gamma (\Delta_\phi-\frac{1-\Delta}{2}  )^2}{\Gamma (1-\frac{\Delta}{2}  )^2 \Gamma (1-\frac{1-\Delta}{2}  )^2}  = \sum\limits_{n=0}^{2\Delta_\phi -2} A_{n} \, \Delta^n (\Delta -1)^n \, ,
\end{align}
where the coefficients $A_n$ can be easily determined case by case in $\Delta_\phi$, and then uses the Casimir equation \eqref{casimir}. The function $\widehat{H}_{0,2}^{B}(w)$ is defined in~\eqref{Hseries}. Just like in the case of the inversion formula, the kernel redefinition~\eqref{Kimpr} introduces at $w=1$ an extra pole, which may spoil the convergence of the integral. As discussed below \eqref{Hboson},  a crossing-symmetric, Regge-bounded subtraction is then required for convergence. Following a similar strategy,  using \eqref{Hunbounded} and \eqref{casimir} we can also derive a dispersion formula for unbounded correlators,
\begin{equation}\label{Kunbounded}
    \begin{split}
        &K^{\text{unbd}}_{\Delta_\phi}(z,w) =K_{\Delta_\phi}(z,w)- \sum_{m,n} A_{m,n} \, \widehat{H}_{m,2}^{B}(w) \,\mathcal{C}^n \left[\frac{2}{\pi^2} \left(\frac{z^2 \log (z)}{1-z}+z \log (1-z)\right)\right] \\
        &- \sum_{m,n} B_{m,n} \, \widehat{H}_{m,1}^{B}(w) \,\mathcal{C}^{n} \left[\frac{2}{\pi^2} \left(\frac{z^2 \log (z)}{1-z}+z \log (1-z)\right)\right] \\
        &- \sum_{m,n} \tilde{A}_{m,n} \, \widehat{H}_{m,2}^{B}(w) \, G_{2+2n}(z) - \sum_{m,n} \tilde{B}_{m,n} \, \widehat{H}_{m,1}^{B}(w) \, G_{2+2n}(z) \, ,
    \end{split} 
\end{equation}
where the coefficients $A_{m,n},B_{m,n},\tilde{A}_{m,n},\tilde{B}_{m,n}$ are fixed as in~\eqref{Hunbounded} (in particular, the last two coefficients are necessary  to improve the behavior of the discrete inversion kernel \eqref{LorInvm}).  Section \ref{sec:pert} discusses explicit examples of this subtraction procedure. 

\subsection{Polyakov blocks in $d=1$}
\label{sec:Polyakov}
The dispersion relation~\eqref{dispersionboson} can be used to compute 1d Polyakov blocks in position space.
As mentioned in the Discussion, Polyakov blocks~\cite{Polyakov:1974gs, Gopakumar:2016wkt}  are crossing symmetric~\footnote{The fact that Polyakov blocks are crossing symmetric implies that their functional form depends on the external dimension $\Delta_\phi$. } and Regge-bounded functions that satisfy~\cite{Mazac:2018mdx,Mazac:2018ycv,Mazac:2018qmi} 
\be
 \mathcal{G}(z) = \sum_{\Delta} a_{\Delta} \mathcal{P}^{\Delta_\phi}_{\Delta}(z)\,.
\ee
Roughly speaking, they are a crossing symmetric version of the conformal blocks \eqref{s-channel}.
Since the four-point function $\mathcal{G}(z)$ can be also expanded in conformal blocks $G_{\Delta} (z)$ with the same coefficients \eqref{s-channel}, the above equation implies
\be\label{polyakovbootstrap}
 \sum_{\Delta} a_{\Delta}  G_{\Delta} (z) =  \sum_{\Delta} a_{\Delta} \mathcal{P}^{\Delta_\phi}_{\Delta}(z)\,.
\ee
Building on the original study of~\cite{Polyakov:1974gs, Gopakumar:2016wkt} in higher-dimensions, eq.~\eqref{polyakovbootstrap} can be turned  into a powerful set of constraints on the OPE data~\cite{Ferrero:2019luz}.  See also~\cite{Ghosh:2021ruh,Kaviraj:2021cvq,Ghosh:2023wjn,Dey:2024ilw} for more recent applications.

Explicitly, the Polyakov block is the crossing-symmetric combination of exchange diagrams in AdS$_2$ in the $s$-, $t$- and $u$-channel,  Regge-improved -- in the bosonic case -- via the subtraction of  the scalar contact diagram of the $\phi^4$ interaction in $AdS_2$\cite{Mazac:2018qmi}. Alternative representations of the Polyakov blocks exist, such as in terms of linear combinations of conformal blocks and their derivatives \cite{Mazac:2018qmi, Ghosh:2023lwe}. Using their representation in terms of master functionals~\cite{Paulos:2020zxx}, they have been computed  in the flat space limit -- the limit of both $\Delta$ and $\Delta_\phi$ large~\footnote{In this limit~\cite{Cordova:2022pbl}, the ratio $\Delta/\Delta_\phi$ is fixed and strictly different from two.} -- in~\cite{Cordova:2022pbl}, where their sum lead to a dispersion relation for the related analytic S-matrix. For specific choices of integer exchanged dimensions $\Delta$ (and $\Delta_\phi=1$) they were computed  in~\cite{Bliard:2022xsm} at tree level, using their explicit definition as sum of exchanged Witten diagrams in AdS$_2$.

The dispersion relation~\eqref{dispersion} (for the fermionic case) or \eqref{dispersionboson} (for the bosonic case) can be used to obtain an integral representation of the Polyakov block, as noticed in \cite{Paulos:2020zxx}.
Indeed, from \eqref{polyakovbootstrap} and the fact that the double discontinuity \eqref{ddisc} commutes with the t-channel OPE, in the bosonic case one has
\begin{equation}
    \dd \big[\mathcal{P}^{\Delta_\phi}_{\Delta}(z)\big] =  2 \sin^{2}{\frac{\pi}{2} (\Delta-2\Delta_{\phi})}\frac{z^{2\Delta_{\phi}}}{(1-z)^{2\Delta_{\phi}}} G_{\Delta}(1-z) \, ,
\end{equation}
whereas in the fermionic one one obtains
\begin{equation}
    \dd \big[\mathcal{P}^{\Delta_\phi}_{\Delta}(z)\big] =  2 \cos^{2}{\frac{\pi}{2} (\Delta-2\Delta_{\phi})}\frac{z^{2\Delta_{\phi}}}{(1-z)^{2\Delta_{\phi}}} G_{\Delta}(1-z) \, .
\end{equation}
Therefore, in the bosonic case and for $\Delta > 2\Delta_\phi$, the Polyakov block can be computed as 
\begin{equation}\label{pb}
   \!\!   \!\!   
       \mathcal{P}^{\Delta_\phi}_{\Delta}(z)\! = 2 \sin^2 ({\textstyle\frac{\pi}{2}(\Delta-2\Delta_\phi))}\!\int_{0}^{1} \!\! \!dw \,w^{-2}\,  K^{\text{bd}}_{\Delta_\phi}(z,w) {\textstyle\frac{w^{2\Delta_\phi}}{(1-w)^{2\Delta_\phi}}}\,(1-w)^\Delta \ _{2}F_{1}(\Delta,\Delta,2\Delta,1-w)\, ,
\end{equation}
with $ K^{\text{bd}}_{\Delta_\phi}(z,w)$ defined in \eqref{Kimpr}.
Notice that the condition $\Delta > 2\Delta_\phi$ is necessary to have a convergent integral~\footnote{This condition has also the effect of killing the extra contour integrals in the bosonic dispersion relation \eqref{dispersionboson}, something in fact necessary as they are defined in terms of the full, unknown, Polyakov block.}.
In the fermionic case instead, one has
\begin{equation}\label{pf}
   \!\!   \!\!   
       \mathcal{P}^{\Delta_\phi}_{\Delta}(z)\! = 2 \cos^2 ({\textstyle\frac{\pi}{2}(\Delta-2\Delta_\phi))}\!\int_{0}^{1} \!\! \!dw \,w^{-2}\,  K_{\Delta_\phi}(z,w) {\textstyle\frac{w^{2\Delta_\phi}}{(1-w)^{2\Delta_\phi}}}\,(1-w)^\Delta \ _{2}F_{1}(\Delta,\Delta,2\Delta,1-w)\,, 
\end{equation}
with $ K_{\Delta_\phi}(z,w)$ defined in \eqref{Kfinal}.
If $\Delta$ is integer, the  integrals above can be computed in closed form, for any fixed $\Delta_\phi$,  and reduce to a combination of polylogarithms and rational functions. This is consistent with the Ansatz discussed in~\cite{Ferrero:2019luz} and with the Witten diagrams computation of~\cite{Bliard:2022xsm}.
 For example, for $\Delta_\phi =1$ and $\Delta=3$ we find
 \begin{equation}
     \begin{split}
         &\mathcal{P}^{\Delta_\phi=1}_{\Delta=3}(z) \textstyle=-\frac{60 z^2 \left(6 z^2-6 z+1\right) \text{Li}_2(1-z) \log (1-z)}{\pi ^2}-\frac{60 \text{Li}_2(z) \left(\left(z^2-6 z+6\right) (z-1)^4 \log (1-z)\right)}{\pi ^2 (z-1)^4 z^2} +\\
         &\textstyle -\frac{60 \text{Li}_2(z) \left((z-2) z \left(6 \left(z^2-z+1\right)^2 \left(2 z^3-z^2-2 z+1\right)+\left(6 z^4-18 z^3+25 z^2-14 z+7\right) z^4 \log (z)\right)\right)}{\pi ^2 (z-1)^4 z^2} + \\
         &\textstyle +\frac{60 \left(6 z^6-6 z^5+z^4-3 z^2+18 z-18\right) \text{Li}_3(1-z)}{\pi ^2 z^2} + \frac{60 z^2 \left(6 z^6-30 z^5+61 z^4-64 z^3+33 z^2-22 z-2\right) \text{Li}_3(z)}{\pi ^2 (z-1)^4} +\\
         &\textstyle-\frac{120 z^2 \left(6 z^2-6 z+1\right) \text{Li}_3\left(\frac{z}{z-1}\right)}{\pi ^2}+\frac{20 z^2 (6 (z-1) z+1) \log ^3(1-z)}{\pi ^2}-\frac{60 \left(z \left(6 z^5-6 z^4+z^3+z-6\right)+6\right) \log ^2(1-z) \log (z)}{\pi ^2 z^2} \\
         &\textstyle +\frac{5 \left(306 z^6-471 z^5+\left(279-62 \pi ^2\right) z^4-153 z^3+219 z^2-144 z+36\right) \log (z)}{\pi ^2 (z-1)^4} +\frac{60 (-192 z^3+183 z^2-90 z+18) \zeta (3)}{\pi ^2 (z-1)^4 z^2}+\\
         &\textstyle-\frac{180 ((z-1) z+1)^2 \left(2 z^4-5 z^3+5 z-2\right) \log (z) \log (1-z)}{\pi ^2 (z-1)^4 z} +\frac{\log(1-z)(-1710 z^7+3075 z^6+5 \left(56 \pi ^2-498\right) z^5+5 \left(306+4 \pi ^2\right) z^4)}{\pi ^2 (z-1)^4 z} \\
         &\textstyle  +\frac{\log(1-z)(-10 \left(249+7 \pi ^2\right) z^3+5 \left(615+4 \pi ^2\right) z^2-1710 z+360)}{\pi ^2 (z-1)^4 z}+ \frac{20 z^5 \left(36 z^2+\pi ^2 (z (6 (z-5) z+61)-62)\right) \tanh ^{-1}(1-2 z)}{\pi ^2 (z-1)^4}+\\
         &\textstyle +\frac{30 \left(((z-3) z+3) \left(2 z^4-z^3+2 z-7\right)+\frac{3 (z-1) ((z-1) z+1)^2}{\pi ^2}+\frac{6}{z}\right)}{(z-1)^3} +\frac{60 (12 z^{10}-60 z^9+122 z^8-128 z^7+78 z^6-38 z^5+113 z^4) \zeta (3)}{\pi ^2 (z-1)^4 z^2}
     \end{split}
 \end{equation}
 For generic $\Delta$  the integral cannot be computed in closed form. It can be however expressed in terms of a combination of infinite series. For instance, for $\Delta_\phi=1$ one has
 \begin{eqnarray}\nonumber
&&\mathcal{P}^{\Delta_\phi=1}_{\Delta}(z) = \textstyle \sin ^2\big(\frac{\pi  \Delta }{2}\big) \Big\{ 2 \csc ^2(\pi  \Delta ) z^{\Delta } \, _2F_1(\Delta ,\Delta ;2 \Delta ;z) \\\nonumber
   &&~~~\textstyle +\frac{2 (-2 \Delta  z ((1-z) \log (1-z)+z \log (z))+\Delta  (z-1) \log (z)+z-1) \, _3F_2(\Delta ,\Delta ,\Delta ;2 \Delta ,\Delta +1;1)}{\pi ^2 \Delta ^2 (z-1)} + \\\nonumber
   &&~~~\textstyle -\frac{4^{\Delta } \Gamma \left(\Delta +\frac{1}{2}\right)}{\pi ^{5/2} (\Delta -2) (\Delta -1) (z-1)^2 \Gamma (\Delta +2)}  \Big( (2-3 (\Delta -1) \Delta ) z (\log(1-z)-\log(z)) + \\\nonumber
   &&~~~\textstyle +\left((\Delta -1) \Delta +((\Delta -1) \Delta -1) z^4-3 (\Delta -1) \Delta  z^3+2 (\Delta -1) \Delta  z^2-1\right) \log (1-z) + \\\nonumber
   &&~~~\textstyle + 2 z^3 \log ((1-z) z)+\left(-2 (\Delta -1) \Delta -3 z^2+2\right) \log (z)\Big) +\\\nonumber
   &&~~~\textstyle -\frac{2 \partial_a \, _3F_2(\Delta ,\Delta ,\Delta;2 \Delta ,\Delta +1+a;1)}{\pi ^2 \Delta } +\frac{2 \partial_a \, _3F_2(\Delta ,\Delta ,\Delta+a;2 \Delta ,\Delta +1;1)}{\pi ^2 \Delta } + \\\nonumber
   &&~~~\textstyle + \sum\limits_{n=0}^{\infty} \Big[\frac{ 2 z^n}{\pi^2} \Big(\frac{z^2 \, _4F_3(\Delta ,\Delta ,n+\Delta ,n+\Delta ;2 \Delta ,n+\Delta +1,n+\Delta +1;1)}{(\Delta +n)^2}-\frac{\log (z) \, _3F_2(\Delta ,\Delta ,n+\Delta -2;2 \Delta ,n+\Delta -1;1)}{\Delta +n-2}\Big) + \\\nonumber
  & &~~~\textstyle +\frac{2z^n}{\pi^2(n-\Delta )^2} \Big( ((n-\Delta ) \log (z)-1) \, _3F_2(\Delta ,\Delta ,\Delta -n;2 \Delta ,-n+\Delta +1;1) +\\\nonumber
   &&~~~\textstyle -(n-\Delta ) \big(\partial_a \, _3F_2(\Delta ,\Delta ,\Delta-n;2 \Delta ,\Delta-n +1+a;1)+ \\\nonumber
   &&~~~\textstyle + \partial_a \, _3F_2(\Delta ,\Delta ,\Delta-n+a;2 \Delta ,\Delta-n +1;1)\big)\Big) + \\\nonumber
   &&~~~\textstyle -\frac{2z}{\pi^2} \Gamma (2 \Delta ) (1-z)^{n-2} \Gamma (n+1) \Big(z^2 \Gamma (\Delta -2) \log \big(\frac{1-z}{z}\big) \, _3\tilde{F}_2(\Delta -2,\Delta ,\Delta ;2 \Delta ,n+\Delta -1;1) \\\nonumber
  & &~~~\textstyle +(z-1)^2 \Gamma (\Delta ) \big(\psi ^{(0)}(\Delta ) \, _3\tilde{F}_2(\Delta ,\Delta ,\Delta ;2 \Delta ,n+\Delta +1;1)+ \\\nonumber
   &&~~~\textstyle + \partial_a \, _3\tilde{F}_2(\Delta ,\Delta ,\Delta;2 \Delta ,\Delta+n +1+a;1) + \partial_a \, _3F_2(\Delta ,\Delta ,\Delta+a;2 \Delta ,\Delta+n +1;1)\big)\Big)\Big]\Big\} \\\label{Pgeneric}
   &&~~~+\frac{z^2}{(1-z)^2}\big(\, z \rightarrow 1-z \,\big) \; .
\end{eqnarray}
where (${}_3\tilde{F}_2$) ${}_3F_2$ are (regularized) generalized hypergeometric functions and  $\partial_a$ indicates the derivative with respect to $a$ evaluated at $a=0$. 
In the fermionic case (half-integer $\Delta_\phi$), the integral~\eqref{pf} can be easily evaluated for integer $\Delta$, as in the bosonic case. For example, for $\Delta_\phi=1/2$ and $\Delta=1$ the Polyakov blocks reads
\begin{eqnarray}   
\mathcal{P}^{\Delta_\phi=1/2}_{\Delta=1}(z) &\textstyle=&-\frac{\left(6 z^2-6\right) \log ^2(1-z) \log (z)}{3 \pi ^2 (z-1)}-\frac{\left(-6 (z-2) z \text{Li}_2(z)-\pi ^2 z^2\right) \log (z)}{3 \pi ^2 (z-1)}+\\\nonumber
       &&\textstyle-\frac{\left(6 \left(z^2-1\right) \text{Li}_2(z)-2 \pi ^2 z^2+2 \pi ^2 z\right) \log (1-z)}{3 \pi ^2 (z-1)}-\frac{18 \left(z^2-1\right) \text{Li}_3(1-z)+\left(18 z^2-36 z\right) \text{Li}_3(z)+18 \zeta (3)}{3 \pi ^2 (z-1)}\,.
\end{eqnarray}

\section{Dispersion relation in perturbation theory}
\label{sec:pert}

A natural domain of application of the inversion formula and the dispersion relation is perturbation theory around a (generalized) free theory. This is because  generalized free theories contain in their spectrum two-particle operators, and their correlators have vanishing double discontinuity, see~ \eqref{ddiscblockbos} and \eqref{ddiscblockfer}. When a small coupling is turned on, the spectrum of these operators receives perturbative corrections and so does the double discontinuity. Nicely, the corresponding contribution to $\dd \big[\mathcal{G}(z) \big]$ at any given order in perturbation theory depends entirely on lower order data. This fact has been heavily exploited in higher dimensions, see for example \cite{Alday:2017vkk,Alday:2018pdi,Alday:2017zzv,Henriksson:2018myn}.  On the other hand, the application of the dispersion relation to perturbation theory presents a few subtleties. First of all, at a given order in perturbation theory it is often the case that  the correlator $\mathcal{G}(z)$ is not Regge bounded. In presence of interactions involving derivatives, one expects the behaviour of the correlator to worsen at every order in the perturbative expansion. At the level of the kernel in the dispersion relation,  this implies that one cannot simply use the expressions \eqref{Kfinal} or \eqref{Kimpr}. Rather,  one has to perform subtractions at every order to match the behaviour of the correlator, as discussed around \eqref{Hunbounded}. The second subtlety is that in general the spectrum of perturbative theories is degenerate, meaning that there are nonequivalent operators with the same quantum numbers at a given perturbative order. 
 As we shall see in concrete cases, this prevents us from computing the double discontinuity order by order using the OPE expansion. 
The degeneracy needs to be solved by studying multiple correlators of non-identical operators~\cite{Aharony:2016dwx,Ferrero:2021bsb,Ferrero:2023gnu}.  Finally, both  kernels  \eqref{Kfinal} and \eqref{Kimpr} are explicitly known only for integer or half-integer external dimensions, meaning that we can only compute four-point functions of protected operators. In the rest of the section we will see two concrete applications: the $\lambda \Phi^4$ theory on the boundary of $AdS_2$ and the defect CFT living on the $1/2$ BPS supersymmetric Wilson line in $\mathcal{N}=4$ Super Yang Mills. While all the correlators that we consider have already been computed using other methods in the literature, here we show how they can be efficiently and elegantly recovered using the dispersion relation, thus showing the power of this tool in the study of one-dimensional CFTs. In the case of the Wilson line, this derivation justifies a posteriori the Ansatz made in \cite{Ferrero:2021bsb}.

\subsection{Massive scalars in $AdS_2$}
\label{sec:scalarsads2}

We start by considering the theory of a massive scalar in $AdS_2$  with the following Lagrangian
\begin{equation}
    \mathcal{L} = \textstyle \frac{1}{2} (\partial \Phi)^2- \frac{m^2}{2}\Phi^2- \frac{\lambda}{4!}\Phi^4 \, .
\end{equation}
The four-point function of the corresponding boundary field $\phi$ with integer dimension $\Delta_\phi$\footnote{We assume we can adjust $m^2$ in such a way that $\Delta_\phi$ is integer, as in \cite{Mazac:2018ycv}.} is
\begin{equation}
     \langle \phi (x_1) \phi(x_2) \phi (x_3) \phi (x_4) \rangle = \frac{1}{(x_{12}\,x_{34})^{2\Delta_\phi}} \,\G(z) \,.
\end{equation}
This correlator was originally computed up to one-loop in $AdS_2$ in \cite{Mazac:2018ycv}  by bootstrapping its OPE data using analytic functionals. With the dispersion relation \eqref{Kimpr} we obtain the same result bypassing a complicated resummation of  OPE data or the need of an Ansatz.

At $\lambda = 0$, the four-point function is the one of a generalized free field (GFF) theory, and can be computed diagrammatically by Wick contractions. The result reads\footnote{In this section we denote with $ \mathcal{G}^{(\ell)}(z)$ the $\ell$th-order correlator in the perturbative expansion. The  zeroth order term should not be confused with  $ \mathcal{G}^{(0)}(z)$ in \eqref{Gextended}.}
\begin{equation}
    \mathcal{G}^{(0)}(z) = 1 + z^{2\Delta_{\phi}} + \frac{z^{2\Delta_{\phi}}}{(1-z)^{2\Delta_{\phi}}} \, .
\end{equation}
The only operators that appear in the conformal block expansion \eqref{s-channel} are two-particle operators $\phi \partial^{2\Delta_{\phi}+2n} \phi$, with OPE data
\begin{equation}\label{GFFdata}
\begin{split}
    &\Delta^{(0)} = 2\Delta_{\phi} + 2n \, ,\\
    &a^{(0)}_n= \frac{2\Gamma^{2}(2\Delta_{\phi} + 2n) \Gamma(4\Delta_{\phi} + 2n - 1)}{\Gamma^{2}(2\Delta_{\phi}) \Gamma(4\Delta_{\phi} + 4n-1) \Gamma(2n +1)} \, .
\end{split}
\end{equation}
We assume the following expansion for the CFT data up to second order
\begin{equation}
\label{pertexp}
    \begin{split}
        &\Delta = 2\Delta_{\phi} + 2n +  \lambda \gamma^{(1)}_n + \lambda^2 \gamma^{(2)}_n+\mathcal{O}(\lambda^3) \, , \\
        &a_{\Delta} = a_{n}^{(0)} +  \lambda\, a_n^{(1)} + \lambda^2 \,a_n^{(2)}+\mathcal{O}(\lambda^3) \, \,,
    \end{split}
\end{equation}
and expand the four-point function $\mathcal{G}(z)$   around the GFF solution
\begin{equation}
    \mathcal{G}(z) = \mathcal{G}^{(0)}(z) + \lambda \, \mathcal{G}^{(1)}(z) +  \lambda^2 \, \mathcal{G}^{(2)}(z)+\mathcal{O}( \lambda^3) \, .
\end{equation}
Comparing the OPE expansion \eqref{s-channel} with  \eqref{pertexp} we find, for the first three orders, 
\begin{eqnarray}\label{pertOPE}
        \mathcal{G}^{(0)}(z) &=&\big(\textstyle\frac{z}{1-z}\big)^{2\Delta_\phi} \sum\limits_n a^{(0)}_n G_{2\Delta_{\phi} +2n}(1-z) \, , \\\label{G1}
        \mathcal{G}^{(1)}(z) &=&\big(\textstyle\frac{z}{1-z}\big)^{2\Delta_\phi} \sum\limits_n \big[ a^{(1)}_n G_{2\Delta_{\phi} +2n}(1-z)+a^{(0)}_n \gamma^{(1)}_n \partial_n G_{2\Delta_{\phi} +2n}(1-z)\big] \, ,\\ \nonumber
        \mathcal{G}^{(2)}(z) &=&\big(\textstyle\frac{z}{1-z}\big)^{2\Delta_\phi} \sum\limits_n \big[ a^{(2)}_n G_{2\Delta_{\phi} +2n}(1-z)+(a^{(0)}_n \gamma^{(2)}_n +a^{(1)}_n \gamma^{(1)}_n) \partial_n G_{2\Delta_{\phi} +2n}(1-z) \, ,\\\label{G2}
        &&~~\qquad\qquad+ \frac{1}{2}a^{(0)}_n (\gamma^{(1)}_n)^2 \partial_n^2 G_{2\Delta_{\phi} +2n}(1-z)  \big] \, .
\end{eqnarray}
Similar expressions hold in the s-channel. Notice that the derivatives of the conformal blocks produce logarithmic terms
\begin{equation}\label{logblock}
   \partial_n G_{2\Delta_{\phi} +2n}(1-z) =  \textstyle \log(1-z) G_{2\Delta_{\phi} +2n}(1-z) + (1-z)^{2\Delta_{\phi} +2n} \partial_n \big[(1-z)^{-2\Delta_{\phi} -2n} G_{2\Delta_{\phi} +2n}(1-z)\big]
\end{equation}
We want to compute the first two orders in the expansion using \eqref{dispersion}. Since the theory does not have derivative interactions, we expect the bound in the Regge limit \eqref{bounded} to hold, therefore we can use the dispersion kernel \eqref{Kimpr}.
The double discontinuity of the correlator, using the expansion~\eqref{pertexp} in~\eqref{ddiscblockbos}, reads then 
\begin{equation}\label{pertddisc}
    \text{dDisc}\big[\mathcal{G}(z)\big] =\lambda^2\, \pi^2 \sum_{n} \frac{1}{2} a^{(0)}_{n} (\gamma^{(1)}_n)^2 \frac{z^{2\Delta_{\phi}}}{(1-z)^{2\Delta_{\phi}}} G_{2\Delta_{\phi} + 2n}(1-z)+\mathcal{O}(\lambda^3) \, .
\end{equation}
In other words the double discontinuity up to second order (one-loop in $AdS_2$) is completely determined by zeroth- and first-order data, a pattern that goes on at higher orders.  This crucial fact is the CFT equivalent of the so-called AdS unitarity, and we refer to \cite{Meltzer:2019nbs} for a thorough discussion of the relation between unitarity cuts in Witten diagrams and the computation of the double discontinuity. By comparing \eqref{pertddisc} with \eqref{G2} and using \eqref{logblock}, one can clearly see that the double discontinuity is completely determined by the terms in \eqref{pertOPE} which are proportional to $\log^n (1-z)$ with $n>1$. Therefore, an equivalent way of determining the double discontinuity is to compute the terms proportional to $\log^n (1-z)$ from the OPE and replace $\log^n (1-z)$ with its double discontinuity \eqref{ddisc}.
As clear from~\eqref{pertddisc}, the double discontinuity at first order (tree-level in AdS) vanishes. This means that the tree-level  correlator results just from the two infinitesimal contour integrals in~\eqref{dispersionboson}
\be \label{disptree}
 \mathcal{G}^{(1)} (z) =  \lim _{\rho \rightarrow 0} \int_{C_\rho^{+}} \frac{dw}{2w^2}  K^{\text{bd}}_{\Delta_\phi=1}(z,w) \mathcal{G}^{(1)}(w)+\lim _{\rho \rightarrow 0} \int_{C_\rho^{-}} \frac{dw}{2w^2} K^{\text{bd}}_{\Delta_\phi=1}(z,w) \mathcal{G}^{(1)}(w) \, ,
\ee
which can be evaluated by expanding both the kernel \eqref{Kfinal} and $\mathcal{G}^{(1)}(w)$ around $w=1$. In particular, the correlator can be expanded using the t-channel OPE~\eqref{G1}. Setting $\Delta_\phi = 1$ for simplicity, we obtain
\begin{eqnarray}
     \mathcal{G}^{(1)} (z)\!\! &= \!\!&\textstyle \lim\limits_{\rho \rightarrow 0} \int_{C_\rho^{+}} dw F(z,w)+  \lim\limits_{\rho \rightarrow 0} \int_{C_\rho^{-}} dw F(z,w)\,,  \\\nonumber 
  F(z,w)\!\!&=\!\!&\textstyle \Big[ \frac{z^2}{\pi^2(1-w)^3} \big(\textstyle \frac{\log(1-z)}{z}\!+\!\frac{\log(z)}{1-z}\big) \!+\!O(\frac{1}{(1-w)^{2}}) \Big] \sum\limits_n \big[ a^{(1)}_n G_{2 +2n}(1-w) +a^{(0)}_n \gamma^{(1)}_n \partial_n G_{2 +2n}(1-w)\big]   
\end{eqnarray}
Switching to radial coordinates $1-w \equiv \rho e^{i \theta}$ and using that $ G_{2 +2n}(1-w) \sim (1-w)^{2+2n}$, the integrals become
\begin{eqnarray}\nonumber
     \mathcal{G}^{(1)} (z) &=&  \lim _{\rho \rightarrow 0} \,z^2\, \big(\textstyle \frac{\log(1-z)}{z}+\frac{\log(z)}{1-z}\big) \int_{0}^\pi d\theta \, \frac{1}{\pi^2 \rho^2} \big[a^{(1)}_0 G_{2}(\rho e^{i \theta}) +a^{(0)}_0 \gamma^{(1)}_0 \partial_n G_{2}( \rho e^{i \theta}) +O(\rho^3)\big] \\\nonumber
 &&+  \lim _{\rho \rightarrow 0} \,z^2\, \big(\textstyle \frac{\log(1-z)}{z}+\frac{\log(z)}{1-z}\big) \int_{2\pi}^{\pi} d\theta \, \frac{1}{\pi^2 \rho^2} \big[a^{(1)}_0 G_{2}(\rho e^{i \theta}) +a^{(0)}_0 \gamma^{(1)}_0 \partial_n G_{2}( \rho e^{i \theta}) +O(\rho^3)\big]\, \\\nonumber
 &=&\lim _{\rho \rightarrow 0} \, \gamma^{(1)}_0 z^2\, \big(\textstyle \frac{\log(1-z)}{z}+\frac{\log(z)}{1-z}\big) \left( \int_{0}^\pi d\theta \, \log( \rho e^{i \theta}) +\int_{2\pi}^\pi d\theta \, \log( \rho e^{i \theta}) \right) \\
 &= &\textstyle 2 \gamma^{(1)}_0 \,z^2\, \big(\frac{\log(1-z)}{z}+\frac{\log(z)}{1-z}\big)\,.
\end{eqnarray}
In the third line, the logarithm comes from the derivative of the conformal block with respect to the conformal dimension and we used the value of $a^{(0)}_0$ in \eqref{GFFdata}. 
All the $n>0$ terms in the OPE expansion of the correlator are suppressed in the limit $\rho \rightarrow 0$,  and 
the only  term contributing is the one proportional to $\gamma_{n=0}^{(1)}$. The latter is an arbitrary constant that can be absorbed in the normalization of the coupling at each order in perturbation theory.  In order to compare with the results in the bootstrap literature, we follow~\cite{Mazac:2018ycv,Ferrero:2019luz} and define our coupling by setting
\begin{equation}\label{couplingdef}
    \gamma^{(1)}_0=1\,,\qquad \quad \gamma^{(\ell)}_0=0 \quad\ \text{for }  \ell>1\,.
\end{equation}
Using this convention we find
\be\label{treephi4}
 \mathcal{G}^{(1)} (z) =2\,z^2\, \Big[\frac{\log(1-z)}{z}+\frac{\log(z)}{1-z}\Big] \,,
\ee
from which one extracts the tree-level OPE data
\be\label{data1}
\gamma^{(1)}_n =\frac{2}{J_{2n+2}} \,,\qquad\qquad
a^{(1)}_n =\frac{1}{2} \partial_n \big[ a^{(0)}_n \gamma^{(1)}_n \big]\,.
\ee
where $J_{\Delta}=\Delta(\Delta-1)$ is the eigenvalue of the $\mathfrak{sl}(2,\mathbb{R})$ Casimir operator. 
At second order, the double discontinuity~\eqref{pertddisc} can be computed from the order zero~\eqref{GFFdata} and order one~\eqref{data1} data, and reads simply
\begin{equation}\label{ddisc2}
        \dd\big[\mathcal{G}^{(2)}(z)\big] =  \frac{\pi^{2}\,z^{2}}{(1-z)^{2}} \log^{2} z \,.
  \end{equation}
To obtain it, the sum in~\eqref{pertddisc} has been computed making use of a standard integral representation for the hypergeometric function defining the conformal block, and exchanging sum and integration. 
Inserting~\eqref{ddisc2} in  the dispersion relation~\eqref{dispersionboson}  one obtains
\begin{equation}\label{G2one}
     \begin{split}
         \mathcal{G}^{(2)}(z) &\stackrel{?}{=}  \int_{0}^{1} \frac{dw}{w^2} K^{\text{bd}}_{\Delta_\phi=1}(z,w) \,  \frac{\pi^{2} w^{2}}{(1-w)^{2}} \log^{2} w +\\
         &+ \lim _{\rho \rightarrow 0} \int_{C_\rho^{+}} \frac{dw}{2w^2}  K^{\text{bd}}_{\Delta_\phi=1}(z,w) \mathcal{G}^{(2)}(w)+\lim _{\rho \rightarrow 0} \int_{C_\rho^{-}} \frac{dw}{2w^2} K^{\text{bd}}_{\Delta_\phi=1}(z,w) \mathcal{G}^{(2)}(w) \, ,
     \end{split}
\end{equation}
However, the first integral does not converge because of a pole in $w=1$. As discussed around \eqref{Kimpr} or \eqref{Hboson}, one may proceed by defining a regularized correlator with  the subtraction 
\begin{equation} \label{subphi4}
    \mathcal{G}^{\text{reg}}(z)=\mathcal{G}^{(2)}(z)-\textstyle\frac{1}{2}\big(z^2 \log^2 (\frac{z}{1-z})+\log^2 (1-z)+\frac{z^2}{(1-z)^2} \log^2 (z) \big)\,.
\end{equation}
In fact, we are free to subtract any function, as long as the resulting regularized correlator is still Regge-bounded and crossing symmetric. 
With the  specific subtraction above, the double discontinuity reads
\be
 \dd\big[\mathcal{G}^{\text{reg}}(z)\big] =  \pi^{2}\big(\textstyle\frac{z^{2}}{(1-z)^{2}} \log^{2} z -\frac{1+z^2}{2}\big)\,,
\ee
and the dispersion relation~\eqref{dispersionboson} is now as a sum of convergent integrals
\be
 \begin{split}
        \mathcal{G}^{\text{reg}} (z) &=  \int_{0}^{1} \frac{dw}{w^2} K^{\text{bd}}_{\Delta_\phi=1}(z,w) \,  \pi^{2}\big(\,\textstyle\frac{w^{2}}{(1-w)^{2}} \log^{2} w -\frac{1+w^2}{2}\big) + \\\label{G2regphi4}
& +\lim _{\rho \rightarrow 0} \int_{C_\rho^{+}} \frac{dw}{2w^2}  K^{\text{bd}}_{\Delta_\phi=1}(z,w) \mathcal{G}^{\text{reg}}(w)+\lim _{\rho \rightarrow 0} \int_{C_\rho^{-}} \frac{dw}{2w^2} K^{\text{bd}}_{\Delta_\phi=1}(z,w) \mathcal{G}^{\text{reg}}(w) \, .
 \end{split}
\ee
For the two semi-circle integrals above, one follows the same strategy as at tree-level, finding
\begin{equation}
         \int_{C_{+}} \frac{dw}{2w^2} K^{\text{bd}}_{\Delta_\phi=1}(z,w) \mathcal{G}^{\text{reg}}(w) +  \int_{C_{-}} \frac{dw}{2w^2} K^{\text{bd}}_{\Delta_\phi=1}(z,w) \mathcal{G}^{\text{reg}}(w) = 
     \textstyle 2 z^2\big(\frac{\log(1-z)}{z}+\frac{\log(z)}{1-z}\big)
\end{equation}
Notice that the result is proportional to the tree-level term~\eqref{treephi4}. 
Computing the integrals in~\eqref{G2regphi4} and using~\eqref{subphi4}, the correlator at second order (one loop in AdS) finally reads
\begin{eqnarray}\nonumber
 &&\mathcal{G}^{(2)}(z)\textstyle
 = \frac{1}{(1-z)^2}\Big[4 (z-2) z^3 \text{Li}_4(1-z)+4  (z^2-1) (1-z)^2 \text{Li}_4(z) \\ \nonumber
 &&\textstyle-2 (1-z)^2 \text{Li}_3(1-z) \Big(\,(z^2-1) \log (1-z)+(z^2+2\,) \log (z)\Big)\\ \nonumber
 &&\textstyle-2 z^2 \text{Li}_3(z)  \big(\,(z^2-2 z+3) \log (1-z)+(z-2) z \log z\,\big)+4 (2 z-1) \text{Li}_4(\frac{z}{z-1}\,) \\\nonumber
  &&\textstyle -\frac{1}{90} \pi ^4 z^2  (z^2-2 z-6)+(\frac{1}{3} \pi ^2 (2 z-1)-(z-1)^2 (z^2+1) \log ^2(z)) \log ^2(1-z)+ \\ \nonumber
 &&\textstyle +(1-z)^2  (\frac{1}{3} \pi ^2 (z^2+2) \log z+z ) \log (1-z)-(z-1) z^2 \log (z) \\ \nonumber
 &&\textstyle+\zeta (3) (2 \log (z)-2 (2 z^3-3 z^2+4 z-1) \log (\frac{z}{1-z}))+\frac{1}{6} (2 z-1) \log ^4(1-z) \\ 
 && \textstyle-\frac{1}{3} (4 z-2) \log (z) \log ^3(1-z)  ]
 \Big]\,,
 \end{eqnarray}
reproducing the result in~\cite{Mazac:2018ycv}~\footnote{See formula (7.34) there.}. One can also extract the CFT data.
A comment here is in order: as often the case in perturbation theory, the spectrum of the operators exchanged in our OPE of $\phi \times \phi$ is degenerate, namely there are many distinct operators with the same dimension at a given order in perturbation theory. Therefore, all the CFT data should be replaced by an average over the degenerate operators, see thorough discussion in~\cite{Ferrero:2021bsb,Ferrero:2023gnu}. At first order we would only be able to extract $\langle \gamma^{(1)}_n \rangle$ instead of $\gamma^{(1)}_n$. If the degeneracy were lifted at first order, we would then not be able to compute the double discontinuity
\begin{equation}\label{G2mix}
    \text{dDisc}\big[\mathcal{G}^{(2)}(z)\big] = \, \pi^2 \sum_{n} \frac{1}{2} \langle a^{(0)}_{n} (\gamma^{(1)}_n)^2 \rangle \frac{z^{2\Delta_{\phi}}}{(1-z)^{2\Delta_{\phi}}} G_{2\Delta_{\phi} + 2n}(1-z) \, ,
\end{equation}
because
\begin{equation}\label{gammamixing}
    \langle a^{(0)}_{n} (\gamma^{(1)}_n)^2 \rangle \neq \langle a^{(0)}_{n} \rangle \langle \gamma^{(1)}_n \rangle^2 \, .
\end{equation}
Thus, to compute   $\langle a^{(0)}_{n} (\gamma^{(1)}_n)^2 \rangle$ in~\eqref{G2mix}, we would need to first solve the degeneracy by properly diagonalizing the dilatation operator. This problem is often referred to as operator mixing. The reason why we ignored this subtlety in this section is that in this particular case the degeneracy is not lifted and one can ignore this problem at least up to one-loop \cite{Mazac:2018ycv}. If one were to compute higher loops, one would need to analyze the spectrum carefully.

\subsection{Defect CFT$_1$ on the 1/2 BPS Wilson line $\mathcal{N}=4$ SYM}
Another interesting one-dimensional CFT is the theory living on the $1/2$ BPS Wilson line in $\mathcal{N}=4$ super Yang-Mills (SYM) with gauge group $SU(N)$. The $1/2$ BPS Wilson line is defined as
\begin{equation}
    \mathcal{W}_{\mathcal{C}}=\frac{1}{N}\text{tr}\,\text{P}\,\text{exp}\int_{\mathcal{C}}\text{d}t\,
\left(i\,A_{\mu}\,\dot{x}^{\mu}+\Phi^6\,|\dot{x}|\right)\,,
\end{equation}
where the contour is a straight line and the trace is taken in the fundamental representation. As usual, $A_\mu$ is the gauge field and $\Phi^i$, with $i=1,...,6$, are the fundamental scalars of  $\mathcal{N}=4$ SYM with $\Delta_\Phi =1$. This line defect preserves the superconformal algebra $\mathfrak{osp}(4^*|4) \supset \mathfrak{sl}(2,\mathbb{R})$.
We are interested in the four-point function
\begin{equation}\label{1111}
    \langle \Phi^i(x_1) \Phi^j(x_2) \Phi^k(x_3) \Phi^l(x_4) \rangle = \frac{\langle \text{tr}\,\text{P}\, \Phi^i(x_1) \Phi^j(x_2) \Phi^k(x_3) \Phi^l(x_4) \mathcal{W}_{\mathcal{C}} \rangle}{\langle \mathcal{W}_{\mathcal{C}} \rangle}= \frac{\G^{ijkl}(z)}{(x_{12}\,x_{34})^{2}} \, ,
\end{equation}
and restrict our attention to  the case of identical operators, e.g. $i=j=k=l=1$. 
Using superconformal Ward identities  \cite{Liendo:2018ukf}, one obtains for $\G^{1111}(z)\equiv  \mathcal{G}(z) $ the structure
\begin{equation}\label{Gward}
   \mathcal{G}(z) = \mathbb{F}(\lambda) \;z^2 +  (2 z^{-1} - 1)f(z) -\left(z^2 - z +1\right)f'(z) \, ,
\end{equation}
where $\mathbb{F}(\lambda)$ is a constant that depends on the   t'Hooft coupling $\lambda=g^2_\text{YM}N$, and $f(z)$ is a crossing-antisymmetric function
 \begin{equation}
            f(z) = - \frac{z^2}{(1-z)^2} f(1-z) \, 
        \end{equation}
 that can be expanded in superconformal blocks~\footnote{We adopt the notation of~\cite{Cavaglia:2022qpg}.}
\begin{equation}\label{superOPE}
f(z) = F_{\mathcal{I}}(z) + { {a_{\mathcal{B}_2} \,  {F}_{\mathcal{B}_2}(z)}}  + \sum_{\Delta } { {a_{\Delta} \,  {F}_{{\Delta}}(z)}} \, .
\end{equation}
Above, $\Delta$ are the dimensions of operators belonging to a long supermultiplet, $a_{\mathcal{B}_2}$ is the squared OPE coefficient of a short operator and the superconformal blocks are 
\begin{equation}
    \begin{split}
       &F_{\mathcal{I}}(z) = z\;,\\
        &F_{\mathcal{B}_2}(z) = z - z\, _2F_1(1,2,4;z ) \;,\\
        &F_{{\Delta}}(z) = \frac{z^{\Delta+1}}{1-\Delta}\, _2F_1(\Delta+1,\Delta+2,2 \Delta+4;z )\,.
    \end{split}
\end{equation}
The constant $ \mathbb{F}(\lambda)$ in~\eqref{Gward} can be computed using supersymmetric localization~\cite{Giombi:2018qox, Liendo:2018ukf}, and at strong coupling reads
\begin{equation}\label{localizationF}
    \mathbb{F}(\lambda) = 1 + a_{\mathcal{B}_2} = 3 - \frac{3}{\lambda^{\frac{1}{2}}}  + \frac{45}{8\lambda^{\frac{3}{2}}} + \frac{45}{4\lambda^{2}} +\mathcal{O}\Big(\textstyle \frac{1}{\lambda^{\frac{5}{2}}}\Big) \, .
\end{equation}
Building on previous work \cite{Giombi:2017cqn,Liendo:2018ukf}, the  large $\lambda$ expansion of the correlator
\begin{equation}\label{Gwilsonpert}
    \mathcal{G}(z) = \mathcal{G}^{(0)}(z) + \frac{1}{\lambda^{\frac{1}{2}}} \, \mathcal{G}^{(1)}(z)+  \frac{1}{\lambda} \, \mathcal{G}^{(2)}(z)+ \frac{1}{\lambda^{\frac{3}{2}}} \, \mathcal{G}^{(3)}(z)+ \frac{1}{\lambda^{2}} \, \mathcal{G}^{(4)}(z) +\mathcal{O}\Big(\textstyle \frac{1}{\lambda^{\frac{5}{2}}}\Big)
\end{equation}
has been recently computed~ \cite{Ferrero:2021bsb,Ferrero:2023gnu} up to fourth order (three loops in $AdS_2$) using analytic bootstrap techniques.
This impressive result was obtained using an Ansatz in terms of a linear combination of Harmonic PolyLogarithms (HPL) multiplied by rational functions, and fixing the unknowns using:
\begin{itemize}
    \item Bose symmetry of the correlator, and in particular crossing symmetry.
    \item The terms proportional to $\log^n (1-z)$ (or $\log^n (z)$)  with $n>1$ in the OPE expansion, which are fixed at each order by lower order data, see for example~\eqref{pertOPE}. In order to compute them one has to take care of operator mixing, especially beyond one-loop.
    \item The assumption\footnote{For the OPE data of the long operators, we define a perturbative expansion as in \eqref{pertexp}.} 
        \begin{equation}\label{gammaregge}
        \gamma^{(\ell)}_n \sim n^{\ell+1} \, ,
    \end{equation}
    on the behaviour of the anomalous dimensions at each perturbative order $\ell$. 
    This corresponds to a divergence $\sim t^\ell$ of the correlator   in the Regge limit~\cite{Ferrero:2019luz}, see~\eqref{ReggeH}.
    \item Compatibility with the OPE \eqref{superOPE}, which combined with the localization result \eqref{localizationF} implies
    \begin{equation}\label{fF}
        f(z) \sim -\frac{\mathbb{F}(\lambda)}{2} z^2 \quad \text{for } z\sim 0 \, ,
    \end{equation}
    There is a similar condition at $z=1$, thanks to crossing.
    This point is essentially equivalent to giving a definition of the coupling.
\end{itemize}
The aim of this section is to reproduce this result from first principles using our dispersion relation, rather than an Ansatz.  
However, we will see that the results obtained from the dispersion relation  depend on undetermined constants, just like in section~\ref{sec:scalarsads2}. We will then need some theory-specific assumptions to fix them, such as the behaviour in the Regge limit and the definition of the coupling. \\ Our strategy will be the following:
\begin{itemize}
    \item Compute the terms proportional to $\log^n (1-z)$ in the OPE expansion \eqref{superOPE} with $n>1$ from lower order data and find $\text{dDisc}\big[\mathcal{G}(z)\big]$, just like we did at one-loop in \eqref{pertddisc}. Regarding the problem of operator mixing, which may prevent us from computing the logarithmic terms from lower order CFT data, we will rely on the analysis of \cite{Ferrero:2023gnu}.
    \item At each perturbative order $\ell$, assume a Regge-behaviour compatible with $\gamma^{(\ell)}_n \sim n^{\ell+1}$ and derive the corresponding unbounded kernel~\eqref{Kunbounded}, using the strategy outlined around~\eqref{Hunbounded}.
    \item Solve the integrals in \eqref{dispersionboson} with the appropriate kernel and regularized correlator. We choose the convenient subtraction
        \begin{equation}\label{Gregpert}
            \mathcal{G}^{\text{reg}}(w) = \mathcal{G}(w) - \sum_{m,n} S_{m,n} \left(\frac{\log ^n(1-z)}{z^m}+\frac{z^2 \log ^n(z)}{(1-z)^{m+2}}\right),
        \end{equation}
    where the coefficients $S_{m,n}$ are fixed demanding that the integrals~\eqref{dispersionboson} do not have singularities at $w=1$. The number of such coefficients depends on the specific case.  Notice that the double discontinuity of the extra term in $\mathcal{G}^{\text{reg}}(z)$ can be easily computed using the definition \eqref{ddisc}, once all the  $S_{m,n}$ are fixed. We stress that~\eqref{Gregpert} is not the only possible subtraction.
    \item Finally, we fix possible undetermined constants using the localization result \eqref{fF}, or equivalently~\footnote{This last equality can be proved by noticing that  \eqref{Gward} implies
    \begin{equation}
     z^{-2} \, \mathcal{G}(z) =   \partial_{z}\left( \mathbb{F}(\lambda) \, z - \Big(\textstyle 1 - \frac{1}{z} +\frac{1}{z^2} \Big) f(z) \right) \, .
    \end{equation}
    Integrating between $z=0$ and $z=1$ and using \eqref{fF} and its equivalent condition at $z=1$ one obtains \eqref{intG}.}
        \begin{equation}\label{intG}
        \int_0^1 \, dz \, z^{-2} \mathcal{G}(z) = 0 \, ,
    \end{equation}
    and, if needed, the following identities for integrated correlators \cite{Cavaglia:2022qpg,Drukker:2022pxk}
    \begin{equation}\label{integratedcorr}
        \begin{split}
          &\int_0^1 dz \, \Big[\Big( \mathcal{G}(z) - \frac{2 (z-1) z+1}{(z-1)^2} \Big) \Big(\frac{1+\log z}{z^2}\Big) \Big]=\frac{3\mathbb{C}(\lambda)-\mathbb{B}(\lambda)}{8 \;\mathbb{B}^2(\lambda)}\; , \\
          & \int_{0}^1 dz \Big[\,  \frac{f(z)}{z} - 2 +\frac{1}{z-1} \Big] = 
        \frac{\mathbb{C}(\lambda)}{4\;\mathbb{B}^2(\lambda)} + \mathbb{F}(\lambda)-3 \; ,
        \end{split}
    \end{equation}
    where $\mathbb{B}(\lambda)$ is the Bremsstrahlung function~\cite{Correa:2012at,Fiol:2012sg} and $\mathbb{C}(\lambda)$ is the curvature function~\cite{Gromov:2015dfa}. Their explicit expressions at large $N$ read
    \begin{eqnarray}
           \mathbb{B}(\lambda) &=&  \frac{\sqrt{\lambda}}{4 \pi^2} \frac{I_2 (\sqrt{\lambda})}{I_1 (\sqrt{\lambda})}  \\
            \mathbb{C}(\lambda) &=&
\frac{\left(2 \pi ^2-3\right) \sqrt{\lambda}}{24 \pi ^4}+\frac{-24 \zeta_3+5-4 \pi
   ^2}{32 \pi ^4}+\frac{11+2 \pi ^2}{64 \pi ^4 \sqrt{\lambda}}+\frac{96 \zeta_3+75+8 \pi ^2}{1024 \pi ^5 \lambda}\\\nonumber
  & &+\frac{3 \left(408 \zeta_3-240
   \zeta_5+213+14 \pi ^2\right)}{16384 \pi ^6 \lambda^{\frac{3}{2}}}+\frac{3 \left(315
   \zeta_3-240 \zeta_5+149+6 \pi ^2\right)}{16384 \pi ^7
   \lambda^2}+O\left(\frac{1}{\lambda^{\frac{5}{2}}}\right)
    \end{eqnarray}
\end{itemize}
 In what follows we will sketch the computation up to three loops. Just like in the $\lambda\Phi^4$ case, we start by computing the order zero term using Wick contractions~\cite{Giombi:2017cqn}
\begin{equation}
    \mathcal{G}^{(0)}(z) = \frac{2 z^2-2 z+1}{(z-1)^2} \, .
\end{equation}
Comparing the above result with \eqref{Gward} and \eqref{superOPE}, one obtains the CFT data for two-particle operators,
\begin{equation}\label{order0data}
          \Delta^{(0)} = 2+2n\,,\qquad\qquad
         \langle a^{(0)}_n \rangle = \frac{\Gamma (5+2n) \Gamma (3+2n) (1+2n)}{\Gamma (6+4n)}\,.
  \end{equation}
Above, we use the average symbol because in this setup operator mixing turns out to be important at higher orders.
Moving on to the tree-level result (first order in the expansion \eqref{Gwilsonpert}), we see that the assumption $\gamma^{(1)}_n \sim n^{2}$ implies that the correlator will diverge linearly in the Regge limit. This means we need to use the unbounded kernel~\eqref{Kunbounded}, which in this case becomes\footnote{We choose to subtract $\widehat{H}_{1,2}^{B}(w)$, but we could have used $\widehat{H}_{0,1}^{B}(w)$ instead. The difference boils down to having a  $\frac{\log(1-w)}{1-w}$ singularity at $w=1$, rather than a second-order pole. We choose the latter for simplicity.}
\begin{eqnarray}\nonumber
     &&     K(z,w) \textstyle=  K_{\Delta_\phi =1}(z,w)- \sum\limits_{n=0}^2 (A_{0,n} \, \widehat{H}_{0,2}^{B}(w) \,+A_{1,n} \, \widehat{H}_{1,2}^{B}(w) ) \mathcal{C}^n \big[\frac{2}{\pi^2} \big(\frac{z^2 \log (z)}{1-z}+z \log (1-z)\big)\big] \\\nonumber
        &&\qquad\textstyle-(\tilde{A}_{0,n} \, \widehat{H}_{0,2}^{B}(w) \,+ \tilde{A}_{1,n} \, \widehat{H}_{1,2}^{B}(w) ) G_{2+2n}(z) = \\\nonumber
        &&\qquad\textstyle= \frac{-2 w^2 \left(2 w^4-9 w^3+16 w^2-14 w+7\right) \left(z^2-z+1\right)^2}{7 \pi ^2 (w-1)^3 (z-1)^2}-\frac{w \left(2 w^4-5 w^3+5 w-2\right) z^2 \left(z^2-z+1\right)^2 \log (1-w)}{\pi ^2 (w (z-1)+1) (w-z) (w (z-1)-z) (w+z-1) \left((w-1) w z^2+z-1\right)}+\\\nonumber
        &&\qquad\textstyle+\Big(-\frac{w^2 \left(w^2-w+1\right) \left(2 w^2-7 w+7\right) z^4}{(w-1)^3}-\frac{w^2 \left(w^2-w+1\right) \left(2 w^2-7 w+7\right) (z-1) z^3}{(w-1)^3} +\frac{7 w^2 (2-z) z^2}{(w-z) (w (-z)+w+z)}+ \\\nonumber
       & &\qquad\textstyle+\frac{w^2 \left(w^2-w+1\right) \left(2 w^2-7 w+7\right) (z-2)}{(w-1)^3}+\frac{2 \left(w^2-w+1\right) \left(2 w^4-7 w^3+14 w-7\right) z^2}{(w-1)^3}
        +\frac{7 (w-1) z^2}{w (z-1)+1}+\frac{7 z^2}{w z-1} \Big) \frac{\log(1-z)}{7\pi^2 z} +\\\nonumber
        &&\qquad\textstyle +\frac{w^4 \left(w^2-w+1\right) z^6 \left(2 w^4 \left(2 z^3-9 z^2+14 z-7\right)+w^3 \left(4 z^4-40 z^3+131 z^2-182 z+91\right)\right) \log (z)}{7 \pi ^2 (w-1)^3 (z-1)^3 (w (z-1)+1) (w+z-1) \left((w-1) w z^2+z-1\right)} +\\\nonumber
        &&\qquad\textstyle +\frac{w^4 \left(w^2-w+1\right) z^6 \left(w^2 \left(-18 z^4+131 z^3-365 z^2+468 z-234\right)+2 w \left(14 z^4-91 z^3+234 z^2-286 z+143\right)\right) \log (z)}{7 \pi ^2 (w-1)^3 (z-1)^3 (w (z-1)+1) (w+z-1) \left((w-1) w z^2+z-1\right)} +\\\label{kerneltree}
        &&\qquad\textstyle +\frac{w^4 \left(w^2-w+1\right) z^6 \left(-14 z^4+91 z^3-234 z^2+286 z-143\right) \log (z)}{7 \pi ^2 (w-1)^3 (z-1)^3 (w (z-1)+1) (w+z-1) \left((w-1) w z^2+z-1\right)}\, ,
\end{eqnarray}
where the coefficients were fixed demanding that $K(z,w)\sim w^{3}$, see~\eqref{ReggeH}.
The discontinuity at this order is zero, just like in the $\lambda \Phi^4$ case. However, since $K(z,w) \sim (1-w)^{-3}$ and $\mathcal{G}(w) \sim (1-w)^0$ from \eqref{superOPE} and \eqref{Gward}, the two contour integrals in \eqref{dispersionboson} diverge because of the singularity at $w=1$. We introduce the regularization \eqref{Gregpert} and fix the coefficients by imposing
\begin{equation}\label{contourvanish}
   w^{-2} K(z,w) \mathcal{G}^{\text{reg}}(w) \sim (1-w)^0 \,\qquad\text{for } w \rightarrow 1\,.
\end{equation}
Since the expansion of $ \mathcal{G}(w)$ around $w=1$ can be read from the t-channel OPE, the condition above fixes the coefficients of the subtraction \eqref{Gregpert} in terms of known and unknown OPE data. In particular, since the block expansion goes schematically like $(1-w)^{2n}$ for two-particle operators, this means that at tree level with $K_{\Delta_\phi}(z,w) \sim (1-w)^{-3}$ the subtraction will depend on $\langle a^{(1)}_0 \rangle, \langle a^{(0)}_0  \gamma^{(1)}_0\rangle,\langle a^{(1)}_1 \rangle, \langle a^{(0)}_1  \gamma^{(1)}_1 \rangle$.
From \eqref{contourvanish}, the contour integrals in \eqref{dispersionboson} vanish
\begin{equation}
     \lim _{\rho \rightarrow 0} \int_{C_\rho^{+}} \frac{dw}{2w^2}  K(z,w) \mathcal{G}^{\text{reg}}(w)+\lim _{\rho \rightarrow 0} \int_{C_\rho^{-}} \frac{dw}{2w^2} K(z,w) \mathcal{G}^{\text{reg}}(w) =0\,.
\end{equation}
The only integral we need to compute is the one involving the double discontinuity of the regularized correlator. The latter can be computed from the explicit form of the subtraction \eqref{Gregpert} and the definition of double discontinuity \eqref{ddisc}, and it will depend on the unknown coefficients $\langle a^{(1)}_0 \rangle, \langle a^{(0)}_0  \gamma^{(1)}_0\rangle,\langle a^{(1)}_1 \rangle, \langle a^{(0)}_1  \gamma^{(1)}_1 \rangle$. 
Solving the integral of the double discontinuity in \eqref{dispersionboson} with $K(z,w)$ defined in \eqref{kerneltree} and removing the subtraction, we obtain
\begin{eqnarray}
   &&     \mathcal{G}^{(1)}(z) = \textstyle -\frac{\left(5  \langle \gamma^{(1)}_1 \rangle (z (2 z-7)+7) z^4+7  \langle \gamma^{(1)}_0 \rangle (z (z (5 z (2 z-7)+49)-28)+14) z^2\right) \log (z)}{245 (z-1)^3}+\\\nonumber
        &&~~\textstyle +\frac{2 (7  \langle \gamma^{(1)}_0 \rangle+ \langle \gamma^{(1)}_1 \rangle) ((z-1) z+1)^2}{49 (z-1)^2}+\frac{\left(5  \langle \gamma^{(1)}_1 \rangle (z (2 z+3)+2) (z-1)^2+7  \langle \gamma^{(1)}_0 \rangle (z (z (5 z (2 z-1)+4)-5)+10)\right) \log (1-z)}{245 z}
\end{eqnarray}
Notice that two of the unknown constants (the OPE coefficients) are automatically canceled once we remove the subtraction. 
Using the theory-dependent constraints \eqref{intG} and one of \eqref{integratedcorr}, we fix the remaining constants and find
\begin{equation}\label{Gtree}
   \mathcal{G}^{(1)}(z)= \textstyle -\frac{2 \left(z^2-z+1\right)^2}{(z-1)^2}+\frac{\left(-2 z^4+z^3+z-2\right) \log (1-z)}{z}+\frac{\left(2 z^4-7 z^3+9 z^2-4 z+2\right) z^2 \log (z)}{(z-1)^3}\,,
\end{equation}
which is precisely the result obtained in \cite{Giombi:2017cqn}.
The only non trivial step of the derivation is the computation of the integral in \eqref{dispersionboson}. While expressions like \eqref{kerneltree} look complicated, an efficient way to do the integrals is the repeated use of
\begin{equation}
    \textstyle \int_0^1 \, dw \, \frac{w^a (1-w)^b}{w-x} =  \textstyle\pi  e^{-i \pi  a} \big(-x^a \csc (\pi  a) (1-x)^b+\frac{(\cot (\pi  a)+i) \Gamma (b+1) \, _2\tilde{F}_1(1,-a-b;1-a;x)}{\Gamma (a+b+1)}\big)\,,
\end{equation}
and its derivatives with respect to parameters $a$ and $b$, to compute the integrals involving logarithms and rational functions.
From \eqref{Gtree}, one can solve \eqref{Gward} for $f(z)$ and then extract the CFT data\cite{Liendo:2018ukf}
\begin{equation}
    \begin{split}
        \langle \gamma^{(1)}_n \rangle = \frac{j^2_{2+2n}}{2}\,, \quad \langle a^{(1)}_n \rangle = \frac{1}{2} \partial_n \langle a^{(0)}_n \gamma^{(1)}_n \rangle\,.
    \end{split}
\end{equation}
where $j^2_\Delta = \Delta(\Delta+3)$ is the eigenvalue of the $\mathfrak{osp}(4^*|4)$ Casimir. 
Moving on to one-loop, one should worry about mixing~\footnote{See the discussion around \eqref{gammamixing}.}. However, the authors of \cite{Ferrero:2021bsb} found  that since, at first order, the anomalous dimension of any operator is proportional to the eigenvalue of the superconformal Casimir, the degeneracy is actually not lifted. Therefore we can compute the higher logarithmic terms and the double discontinuity at one-loop from tree level data, just like we did in \eqref{pertddisc}. The double discontinuity in this case reads
\begin{equation}
    \dd \big[\mathcal{G}^{(2)}(z)\big]=\frac{\pi ^2 \left(9 z^6-8 z^5+4 z^4+4 z^2-8 z+9\right)}{2 z^2}\,.
\end{equation}
Assuming  $\gamma^{(2)}_n \sim n^{3}$,  we see that the correlator diverges as $t^2$ in the Regge limit. We can still use the same kernel as for the tree level \eqref{kerneltree}, since it goes like $\mathcal{O}(w^4)$ for small $w$. We fix the coefficients in the subtraction \eqref{Gregpert} by demanding that the integral of the double discontinuity in \eqref{dispersionboson} converges and that $w^{-2} \, K(z,w) \mathcal{G}^{\text{reg}}(w) \sim (1-w)^0$, killing the contour integrals. Performing the remaining integral and removing the subtraction, we find
\begin{equation}
     \begin{split}
        &\mathcal{G}^{(2)}(z) = \textstyle \frac{z^2 (9 z^6-46 z^5+99 z^4-116 z^3+83 z^2-30 z+10) \log ^2(z)}{2 (z-1)^4}+\frac{\left(z \left((z (9 z-8)+4) z^3+4 z-8\right)+9\right) \log ^2(1-z)}{2 z^2} \\
        &\textstyle +\frac{(118 z^6+125 z^5+4388 z^4-10102 z^3+4388 z^2+125 z+118)\log (1-z)}{840 (z-1)^2 z} +  \frac{\left(-118 z^6+833 z^5-6783 z^4\right) \log (z)}{840 (z-1)^3}  \\
        &\textstyle -\frac{(3 (z-1) z+1) \left(6 z^6-18 z^5+19 z^4-8 z^3+z-2\right) \log (z) \log (1-z)}{2 (z-1)^3 z} +\frac{\left(11060 z^3-3430 z^2-2520 z+840\right) \log (z)}{840 (z-1)^3}  \\
        &\textstyle +\langle \gamma^{(2)}_0 \rangle \Big[-\frac{(z (z (5 z (2 z-7)+49)-28)+14) z^2 \log (z)}{35 (z-1)^3}+\frac{2 ((z-1) z+1)^2}{7 (z-1)^2}+\frac{(z (z (5 z (2 z-1)+4)-5)+10) \log (1-z)}{35 z}\Big] \\
        &\textstyle +\langle \gamma^{(2)}_1 \rangle \Big[-\frac{(z (2 z-7)+7) z^4 \log (z)}{49 (z-1)^3}+\frac{2 ((z-1) z+1)^2}{49 (z-1)^2}+\frac{(z-1)^2 (z (2 z+3)+2) \log (1-z)}{49 z}\Big] -\frac{1831 ((z-1) z+1)^2}{420 (z-1)^2} \, .
    \end{split}
\end{equation}
Using \eqref{intG} and one of the integrated correlators in \eqref{integratedcorr}, we obtain \cite{Liendo:2018ukf}
\begin{eqnarray}\nonumber
        &&\mathcal{G}^{(2)}(z) = \textstyle \frac{\left(9 z^6-8 z^5+4 z^4+4 z^2-8 z+9\right) \log ^2(1-z)}{2 z^2}+\frac{(26 z^6-63 z^5+66 z^4-62 z^3+66 z^2-63 z+26) \log (1-z)}{4 (z-1)^2 z}\\\nonumber
        &&~~\textstyle +\frac{z^2 \left(9 z^6-46 z^5+99 z^4-116 z^3+83 z^2-30 z+10\right) \log ^2(z)}{2 (z-1)^4} +\frac{\left(-26 z^6+93 z^5-141 z^4+92 z^3-36 z^2-12 z+4\right) \log (z)}{4 (z-1)^3}\\
        &&~~\textstyle +\frac{\left(-18 z^8+72 z^7-117 z^6+99 z^5-43 z^4+5 z^3+9 z^2-7 z+2\right) \log (z)  \log (1-z)}{2 (z-1)^3 z} +\frac{2 \left(z^2-z+1\right)^2}{(z-1)^2} \, .
\end{eqnarray}
From which one can extract the same CFT data that was found in~\cite{Ferrero:2023gnu}
\begin{eqnarray}\nonumber
&&\langle\gamma^{(2)}\rangle_{n}=\textstyle\gamma^{(1)}_{n}\,\frac{1}{2} \partial_{n}\gamma^{(1)}_{n}+\frac{j^2_{2+2n}}{8}\big(-11-\frac{6}{j^2_{2+2n}+2}+4\,H_{3+2n}\big)\,,\\ 
&&\langle a^{(2)} \rangle_{n}=\frac{1}{2} \partial_{n} \langle a^{(0)}\,\gamma^{(2)}+a^{(1)}\,\gamma^{(1)}\rangle_{n}-\frac{1}{4}\partial_{n}^2\langle a^{(0)}\,(\gamma^{(1)})^2 \rangle_{n}\\\nonumber
&&\textstyle+\langle a^{(0)} \rangle_{n}\,\big(\frac{j^2_{2+2n}(j^2_{2+2n}-2)}{2}(S_{-2}(2+2n)+\frac{1}{2}\zeta(2))-\textstyle \frac{1068 + 5000 n + 8772 n^2 + 7616 n^3 + 3424 n^4 + 736 n^5 + 64 n^6}{4(j^2_{2+2n}+2)}\big)\,,
\end{eqnarray}
 and
\begin{eqnarray}
H^{(m)}_n&=&\sum_{k=1}^n\frac{1}{k^m}\,,  \qquad\qquad
H_n\equiv H^{(1)}_n\,,\\
S_{-2}(n)&=&\sum_{k=1}^n\frac{(-1)^k}{k^2}=\frac{(-1)^n}{4}\left(H^{(2)}_{n/2}-H^{(2)}_{(n-1)/2}\right)-\frac{1}{2}\zeta(2)\,.
\end{eqnarray}
The procedure can be carried on at the next two orders,  we will however skip the details of the computations and highlight the differences with respect to the previous orders. First of all, in order to compute the double discontinuity at three loops one needs $\langle a^{(0)}\,(\gamma^{(2)})^2\rangle_{n}$. It turns out that at this order the degeneracy is lifted, therefore
    \begin{align}
        \langle a^{(0)}\,(\gamma^{(2)})^2\rangle_{n} \neq 
        \langle a^{(0)}\rangle_{n}\,\langle\gamma^{(2)}\rangle_{n}^2\,,
    \end{align}
and one has to solve the operator mixing problem in order to compute the double discontinuity. This was done in \cite{Ferrero:2021bsb} and we report their result
\begin{eqnarray}\label{gamma2squared}
\frac{\langle a^{(0)}\,(\gamma^{(2)})^2\rangle_{n}}{\langle a^{(0)}\rangle_{n}}&=&\langle \gamma^{(2)}\rangle_{n}^2+\frac{1}{2} j^2_{2+2n} (j^2_{2+2n}-2)S_{-2}(3+2n)+\frac{1}{8}j^2_{2+2n} (3j^2_{2+2n}-4)H_{3+2n}^2\\\nonumber
&&+\textstyle\Big(-j^4_{n}+\frac{3}{4}\big(5+\frac{2}{j^2_{2+2n}+2}\big)\Big)H_{3+2n}+\frac{1}{32}\Big(-156+50 j^2_{2+2n} +29 j^4_{n}+\frac{24}{j^2_{2+2n}+2}\Big)\,.
\end{eqnarray}
Using this result, together with the other OPE data from previous orders, we can compute all the higher logarithmic terms in the OPE at two and three loops. The corresponding contributions to the double discontinuity read respectively
\begin{eqnarray}\nonumber
\text{dDisc}[\mathcal{G}^{(3)}(z)]&=& -\big[111 z^6-106 z^5+80 z^4+80 z^2-106 z+111\big]\textstyle\frac{\pi^2}{4 z^2} \\
&+& \big[\!-\!72 z^8+99 z^7-68 z^6+16 z^5+16 z^3-68 z^2+99 z-72\big]\textstyle\frac{\pi ^2\log (1-z)}{2 z^3} \nonumber\\
&+& \big[72 z^7-108 z^6+76 z^5-20 z^4+4 z^2-8 z+9\big]\textstyle\frac{\pi ^2\log z}{2 z^2}  
\end{eqnarray}
and
\begin{eqnarray} \nonumber
&&\!\!\!\!\!\!\!\text{dDisc}[\mathcal{G}^{(4)}(z)]\!=\!\big[\! -\! 900 z^{10}\!+\!1728 z^9\!-\!1539 z^8\!+\!668 z^7\!-\!82 z^6\!-\!82 z^4\!+\!668 z^3-\!1539 z^2\!\!\\\nonumber
&&+\!1728 z\!-\!900\big]\textstyle\frac{ \pi ^4 }{24 z^4}+\big[5184 z^{10}\!-\!16134 z^9\!+\!21705 z^8\!-\!16464 z^7\!+\!11481 z^6\!-\!16464 z^5\!\\\nonumber
&&+\!21705 z^4\!-\!16134 z^3\!+\!5184 z^2 \big]\textstyle\frac{\pi ^2}{48 z^4 (z-1)^2}-\big[ 6048 z^{10}\!-\!27450 z^9 \!+\!53586 z^8\!-\!58811 z^7\!\\\nonumber
&&+\!38769 z^6\!-\!14296 z^5\!+\!1636 z^4\!+\!1407 z^3  -\!1417 z^2\!+\!768 z\!-\!180\,\big]\,\textstyle\frac{\pi^2\,\log (z)}{16(z-1)^3 z^2}\nonumber\\
&&+\big[ 3024 z^{10}-10611 z^9+15798 z^8-12899 z^7+5782 z^6  -2184 z^5+5782 z^4-12899 z^3\nonumber\\
&&+15798 z^2-10611 z+3024 \big]\,\textstyle\frac{\pi ^2  \log (1\!-\!z)}{8 (z-1)^2 z^3} + \big[900 z^{10}\!-\!1728 z^9\!+\!1539 z^8\!-\!668 z^7\!+\!82 z^6\!
+\!82 z^4\!
\nonumber\\\nonumber
&&-\!668 z^3 +1539 z^2-1728 z+900 \big] \textstyle\frac{\pi^2\log ^2(1\!-\!z)}{4 z^4} +\big[3600 z^{12}-21888 z^{11}+58572 z^{10}-90696 z^9
\\\nonumber
&&+89316 z^8-57264 z^7+23176 z^6   -5484 z^5+1118 z^4-932 z^3+828 z^2-400 z+81 \big]\textstyle \frac{\pi ^2  \log ^2(z)}{16 (z-1)^4 z^2} \\
&&-\big[1800 z^{12}-9072 z^{11}+19791 z^{10}-24481 z^9+18629 z^8-8631 z^7+2129 z^6+137 z^5-1053 z^4\nonumber\\
&& +1767 z^3 -1743 z^2+945 z-216 \big]\textstyle\frac{\pi ^2  \log (z) \log (1\!-\!z)}{4 (z-1)^3 z^3}+\big[96 z^{11}\!-\!423 z^{10}\!+\!785 z^9\!-\!799 z^8\nonumber\\
&& +477 z^7\!-\!137 z^6\!-\!137 z^5\!+\!477 z^4\!-\!799 z^3\!+\!785 z^2\!-\!423 z\!+\!96 \big]\textstyle\frac{3\pi^2 \text{Li}_2(1\!-\!z)}{8 (z-1)^3 z^3}
\end{eqnarray}
A second new aspect of the calculation at these orders is that, since the behaviour in the Regge limit gets worse, see~\eqref{gammaregge},  one is forced to subtract an extra term from the dispersion kernel~\footnote{As it happened at previous orders, at fourth order one can use the kernel used for the third order calculation. }
\begin{eqnarray}\nonumber
        K(z,w) &&=\textstyle K_{\Delta_\phi=1}(z,w)- \sum_{m=0}^3 \sum_{n} A_{m,n} \, \widehat{H}_{m,2}^{B}(w) \,\mathcal{C}^n \left[\frac{2}{\pi^2} \left(\frac{z^2 \log (z)}{1-z}+z \log (1-z)\right)\right] \\
        &&\textstyle- \sum_{m=0}^3 \sum_{n} \tilde{A}_{m,n} \, \widehat{H}_{m,2}^{B}(w) \, G_{2+2n}(z) \, ,
\end{eqnarray}
where the coefficients are fixed demanding that $K(z,w)\sim w^6$ for small $w$, according to~\eqref{ReggeH}. We do not report them here to avoid cluttering.  In this case $K(z,w)\sim(1-w)^{-5}$~for~$w\sim1$, and demanding  the convergence of the integrals in the dispersion relation~\eqref{dispersionboson}  introduces a dependence on extra, unknown OPE data. This means that,  to fix these unknown constants,  one has to impose an additional constraint, so that one can use equation~\eqref{intG} together with both the integrated correlators~\eqref{integratedcorr}. For the third and fourth-order four-point function~\eqref{1111}, we checked that the result obtained using the dispersion relation~\eqref{dispersionboson} reproduces the ones in~\cite{Ferrero:2023gnu}.

\section*{Acknowledgments} 
We are grateful to L. Bianchi, G.~Bliard and G.~Peveri for useful discussions.
The research of DB and VF received partial support through the STFC grant ST/S005803/1. The
research of VF is supported by the DFG Heisenberg Professorship program 506208580.

\bibliographystyle{nb}
\bibliography{references-unitarity}

\end{document}